\RequirePackage{fix-cm}
\documentclass{article} 

\usepackage{graphicx}
\usepackage{epstopdf}
\usepackage{xcolor}

\usepackage{amsmath}
\usepackage{mathtools} 
\usepackage{amssymb}

\usepackage{booktabs} 
\usepackage{csquotes}
\usepackage{microtype}

\usepackage[]{caption}
\usepackage{subcaption}

\usepackage{adjustbox}

\usepackage{tikz}
\usetikzlibrary{external,patterns,shapes,arrows,calc,positioning,fit,spy}

\usepackage[%
bibencoding=utf8,
giveninits=true, 
style=numeric-comp, 
clearlang=true, 
hyperref=true,
doi=false, 
natbib=true, 
isbn=false, 
url=false, 
eprint=false,
maxbibnames=20, 
date=short, 
sortcites,
sorting=nty,
date=year,
backend=bibtex
]{biblatex}
\bibliography{references}

\usepackage{pgfplots}
\pgfplotsset{compat=newest}

\usepackage{standalone}

\input{commands.tex}

\definecolor{freeflow}{HTML}{acd7f3}
\colorlet{porousmedia}{brown!50!white}
\definecolor{blue1}{RGB}{0,102,189}
\definecolor{blue2}{RGB}{98,160,214}
\definecolor{my_orange}{RGB}{243,98,33}

\tikzstyle{flowSolver} = [rectangle, draw, fill=blue1!60, 
text width=2em, text centered, rounded corners, minimum height=2em]
\tikzstyle{solidSolver} = [rectangle, draw, fill=brown!70, 
text width=2em, text centered, rounded corners, minimum height=2em]
\tikzstyle{accelerator} = [rectangle, draw, fill=my_orange!60, 
text width=2em, text centered, rounded corners, minimum height=2em]
\tikzstyle{coord} = [coordinate]
\tikzstyle{line} = [draw, -latex']

\pgfplotsset{
  reference-solution/.style={
    black, every mark/.append style={solid, fill=black},mark=square*, thick,mark repeat={50}
  },
  precice-solution/.style={
    red, every mark/.append style={solid, fill=red},mark=diamond*, thick,mark repeat={50}
  },
  precice-solution-imvj/.style={
    blue, dashed, every mark/.append style={solid, fill=red},mark=diamond*, thick,mark repeat={50}
  },
  flowPressurePlotStyle/.style={
    xlabel = {$x^{\fractureDomain}$ [m]},
    ylabel = {$\hat{p}$ [kPA]},
    grid=major,
    xmin=5.14,
    xmax=6.25,
    ymin=-20000,
    ymax=2e5,
    xtick distance=0.2,
    xticklabels={-0.0,0.1,0.2, 0.4, 0.6,0.8,1.0,1.2},
    legend style={
      name=legend,
      at={(0.98,0.98)},
      anchor=north east,
      font=\scriptsize,
    },
    legend cell align={left},
    thick,
    tick label style={font=\scriptsize},
    label style={font=\footnotesize},
    width=\textwidth,
    height=0.3\textwidth,
    scale only axis=true,
  },
  couplingIterationsPlotStyle/.style={
    xlabel = {Time step},
    ylabel = {Coupling iterations in time step},
    grid=major,
    xmin=1,
    xmax=20,
    ymin=0,
    ymax=40,
    legend style={
      name=legend,
      at={(0.98,0.98)},
      anchor=north east,
      font=\tiny,
    },
    thick,
    tick label style={font=\tiny},
    label style={font=\scriptsize},
    width=\textwidth,
    height=0.3\textwidth,
    scale only axis=true,
  },
}

\usepackage[affil-it]{authblk}

\newcommand*{\fixedPointProblem}{\mathit{H}}
\newcommand*{\fixedPointProblemFunc}[1]{\fixedPointProblem(#1)}
\newcommand*{\fixedPointProblemFuncInv}[1]{\fixedPointProblem^{-1}(#1)}

\newcommand*{\svdTrunctationThreshold}{\varepsilon_{\mathrm{SVD}}}
\newcommand*{\filterLimit}{\varepsilon_{\mathrm{F}}}

\newcommand{\Vi}{\ensuremath{V_i}}
\newcommand{\Wi}{\ensuremath{W_i}}
\newcommand{\Vin}{\ensuremath{V^n_i}}
\newcommand{\Win}{\ensuremath{W^n_i}}
\newcommand{\timeReuseParameter}{\ensuremath{m}}
\newcommand{\iterationReuseParameter}{\ensuremath{M}}
\newcommand*{\relaxationParameter}{\omega}
\newcommand*{\imvjRestart}{\ensuremath{\eta}}

\author[1]{%
  \large Patrick Schmidt%
  \thanks{\texttt{patrick.schmidt@mechbau.uni-stuttgart.de}}%
}

\author[2]{%
  \large Alexander Jaust%
  \thanks{\texttt{alexander.jaust@ipvs.uni-stuttgart.de}}%
}

\author[1]{%
  \large Holger Steeb%
  \thanks{\texttt{holger.steeb@mechbau.uni-stuttgart.de}}%
}

\author[2]{%
  \large Miriam Schulte%
  \thanks{\texttt{miriam.schulte@ipvs.uni-stuttgart.de}}%
}

\affil[1]{\normalsize Institute of Applied Mechanics (CE), University of Stuttgart, Pfaffenwaldring~7,  D-70\,569 Stuttgart, Germany}

\affil[2]{\normalsize  Institute for Parallel and Distributed Systems, University of Stuttgart,  Universit\"{a}tsstra{\ss}e~38, D-70\,569 Stuttgart, Germany}

\begin{document}

\title{\bf Simulation of flow in deformable fractures using a quasi-Newton based partitioned coupling approach}

\maketitle

\begin{abstract}
We introduce a partitioned coupling approach for iterative coupling of flow processes in deformable fractures embedded in a poro-elastic medium that is enhanced by interface quasi-Newton (IQN) methods. In this scope, a unique computational decomposition into a fracture flow and a poro-elastic domain is developed, where communication and numerical coupling of the individual solvers are realized by consulting the open-source library \preCICE. The underlying physical problem is introduced by a brief derivation of the governing equations and interface conditions of fracture flow and poro-elastic domain followed by a detailed discussion of the partitioned coupling scheme. We evaluate the proposed implementation and undertake a convergence study to compare a classical interface quasi-Newton inverse least-squares (IQN-ILS) with the more advanced interface quasi-Newton inverse multi-vector Jacobian (IQN-IMVJ) method. These coupling approaches are verified for an academic test case before the generality of the proposed strategy is demonstrated by simulations of two complex fracture networks. In contrast to the development of specific solvers, we promote the simplicity and computational efficiency of the proposed partitioned coupling approach using \preCICE and \FEniCS for parallel  computations of hydro-mechanical processes in complex, three-dimensional fracture networks.
\end{abstract}

\section{Introduction}
\label{sec:intro}
Modeling of transient flow processes in deformable high-aspect ratio fractures (length $\gg$ aperture) embedded in a poro-elastic medium is a non-trivial task. Complex discretization of the fracture(-network) geometry and stiff numerical coupling of the surrounding poro-elastic and fracture-flow domain require specific implementations to guarantee stability and efficiency of the designed solver. In this work, we propose a highly efficient, straightforward implementation of the governing partial differential equations (PDEs) of each domain using the open-source package \FEniCS \cite{fenics} and realizing the numerical coupling process, its acceleration via quasi-Newton methods and inter-solver communication by the open-source library \preCICE{} \cite{preCICE}.

Phenomena such as the occurrence of reverse water-level fluctuations in distant monitoring wells during hydraulic testing of fractured reservoirs indicate the importance of hydro-mechanical interaction throughout fracture-flow processes \cite{gellasch2013,slack2013}. Perturbations of the reservoir's equilibrium state result in immediate non-local fracture deformations and time delayed pressure diffusion with distinct characteristic time scales, which highly affect the measured pressure transients. Once the fracture deformation influences the pressure evolution, traditional \cite{muskat1938flow,fetter2001} and extended \cite{ortiz2011} diffusion-based models fail and the necessity of consistent hydro-mechanical simulations arises \cite{vinci2014}. Creeping flow conditions in high-aspect ratio fractures motivate to simplify the balance equations based on pressure-driven, Poiseuille-type flow formulations \cite{louis1969,witherspoon1980} to capture flow processes in deformable fractures. The resulting fracture-flow equation leads to a reduction of space dimension of the computational domain by one. Equivalently, the numerical implementation is introduced as a hybrid-dimensional model \cite{vinci2014,vinci2015}, which can be simplified to the lubrication equation \cite{batchelor2000} by dimensional analysis considering the specific case of high-aspect ratio geometries \cite{vinci2014}.

The numerical approaches addressed in literature to solve the strongly coupled hydro-mechanical fracture-flow problem can be divided in two groups: (i) monolithic and (ii) partitioned schemes. Monolithic approaches introduce the fracture-flow domain by zero-thickness interface elements \cite{segura2008I,segura2008,Jin2017,settgast2016} and mostly require direct solution strategies to guarantee high numerical stability since distinct characteristic properties of both domains lead to poor conditioning of the global system \cite{Schmidt2019}. In general,  tailored meshing and integration strategies are required to discretize the fracture domain by interface elements. The complexity dramatically increases in the presence of intersecting fractures or regions containing fracture tips. In contrast, partitioned schemes allow the individual treatment of the subdomain and, in particular, the use of non-conformal meshes \cite{Schmidt2019}. However, iterative coupling of both domains is required to ensure global equilibrium for each conducted time step, which potentially makes this approach computationally more expensive. Therefore, stabilization and acceleration methods are used to keep the number of coupling iterations needed low. In the literature, numerical convergence and stability of the equilibrium iterations have been achieved by adapting physics-based preconditioning of unfractured \cite{kim2011I,kim2011II} to the specific case of fractured poro-elastic media \cite{girault2015,girault2016,castelletto2015,keilegavlen2017,berge2020}. Nevertheless, parallel communication within the proposed staggered algorithms is nontrivial, but required to solve large systems in reasonable time. Therefore, we propose an easily accessible, efficient and fully parallelized deformable fracture-flow implementation introducing a unique computational split of the fracture-flow and poro-elastic domain. It uses advanced quasi-Newton schemes for stabilization and acceleration of the coupling.

More specifically, we employ two separate solvers for (i) flow and mechanical deformation of the
porous structure and (ii) flow in the fracture itself. This allows for the use of available, highly efficient and parallel solvers for each of these two subdomains. We avoid assembling the ill-conditioned monolithic system of equations. The coupling library \preCICE{} \cite{preCICE} is used to establish and steer the iterative coupling between the two solvers. \preCICE{} provides a generic coupling solution interface, that is not tailored to a specific use case. Its library approach makes it simple to use with a large variety of solvers. It comes with adapter codes for popular simulation software and a high-level interface for many popular programming languages. Additionally, it has been optimized for high computational efficiency by implementing several acceleration methods, including a number of interface quasi-Newton methods \cite{Scheufele2017}, parallel data communication \cite{Lindner2019,Lindner2020}, and sophisticated data mapping techniques including radial basis function mapping \cite{Lindner2019,Lindner2017} for the use with non-matching grids. This sets \preCICE apart from commercial or closed sourced alternatives such as MpCCI\footnote{www.mpcci.de} \cite{Joppich2006}, where one cannot access or change the implementation. Other open-source alternatives such as ADVENTURE\footnote{https://adventure.sys.t.u-tokyo.ac.jp/} \cite{Kataoka2014}, the Data Transfer Kit\footnote{https://ornl-cees.github.io/DataTransferKit/} \cite{Slattery2013} or OpenPALM\footnote{https://www.cerfacs.fr/globc/PALM\_WEB/} \cite{Buis2006}, e.g., are either lacking the high-level programming interface or combined implementation of highly-parallel data communication, coupling steering, sophisticated data mapping, and coupling acceleration. Combining the efficient iterative solvers in the subdomains with \preCICE to realize the iterative coupling of both domains yields a fast and efficient overall solution technique. It eliminates the need to develop a specialized solver or preconditioner, respectively, for the corresponding monolithic system. Thus, our simulations of hydro-mechanical processes at high resolution and accuracy can be performed by simple high-level solvers written in Python based on the finite-element library \FEniCS. We use the Python interface of \preCICE since the recently developed \FEniCS-\preCICE adapter~\cite{Rodenberg2021} had not been available and was lacking some functionality needed for the applications presented during the preparation of our work.

After briefly deriving the governing equations, see Sec.~\ref{sec:eqs}, and introducing the coupling scheme, see Sec.~\ref{sec:couplig}, we verify and investigate the proposed coupling approach for an academic test case (single fracture embedded in a three-dimensional poro-elastic domain) at high resolution. The results obtained by the staggered scheme are compared to a monolithic approach to proof consistency of both strategies in Sec.~\ref{sec:results_verification}. Afterwards, we conduct a case study investigating the convergence behavior of different interface quasi-Newton schemes and their dependency on coupling specific parameters, see Sec.~\ref{sec:results_parallel}. We demonstrate the ability of this approach for large problems for three different meshes ranging from tens of thousands to several million degrees of freedom (DoF), see Sec.~\ref{sec:convergence:meshes}. Subsequently, the relevance of the proposed method for more complex applications is shown by two more challenging test cases. The first case focuses on flow processes in complex fracture networks on a short (minutes) time-scale, see Sec.~\ref{sec:results_network}, while, in the second case, flow through fractured porous media on a large time-scale (days up to years) is investigated, see Sec.~\ref{sec:results_porous}. In summary, this work introduces a stable implementation for fully coupled, parallelized three-dimensional simulations of flow processes in deformable fractures, using an unique split of the computational fracture-flow and poro-elastic domain.

\section{Governing equations}
\label{sec:eqs}
We introduce the governing equations for both parts of our fracture system, the fracture(s) $\fractureDomain$ and the surrounding poro-elastic $\poroElasticDomain$ domain, resulting in a coupled formulation capturing solid deformation and fluid flow in fractured porous media.
For the fracture domain, we base our derivation on the observation that (i) deformation induced volume changes of high-aspect ratio fractures have a strong impact on the flow solution and require an implicit coupling to the surrounding solid domain; (ii) explicit three-dimensional modeling of flow processes in such fracture systems is challenging and leads to poor results in case of insufficient mesh quality.

We overcome these challenges by lower dimensional modeling of the fracture flow domain where the fracture aperture is implicitly treated as a function on a two-dimensional manifold. The value of this function is given by an initial opening value and the poro-elastic deformation of the surrounding porous medium. 

\subsection{Flow in a deformable fracture $\fractureDomain$}
In the following, we derive a hybrid-dimensional model \cite{vinci2014,vinci2015} governing the flow processes in deformable high-aspect ratio fractures by evaluation of the balance of mass and momentum within the fracture domain $\fractureDomain$. Equilibrium conditions of the fluid and the biphasic porous medium are enforced in terms of fracture aperture $\aperture$ and fluid pressure $p$.

\begin{figure}[htb]
\centering
\input{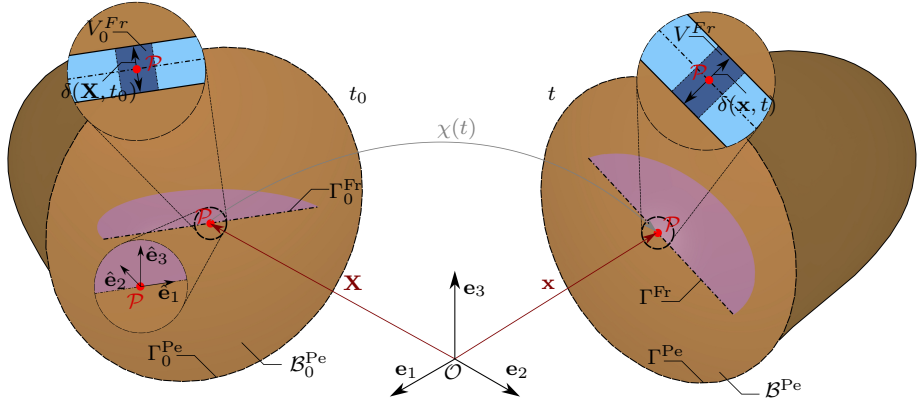}
\caption{Three dimensional representation of an embedded fracture characterized by its surface $\fractureSurface$ in a deformable poro-elastic body $\poroElasticDomain$ with its boundary $\poroElasticSurface$. A local coordinate system $\hat{\mathbf{e}}_i$ at the fracture level is introduced, where $\hat{\mathbf{e}}_3$ is pointing in the direction of the fracture surface normal $\mathbf{n}^\text{Fr}$. The volume $V^\text{Fr}$ and aperture $\aperture$ changes of the fracture between the reference configuration $\mathbf{X}$ at time $t_0$ and the current configuration $\coordinateVec$ at time $t$ of a material point $\mathcal{P}$ are implicitly coupled to the poro-elastic deformations expressed by the unique motion function $\mathbf{\mathcal{\chi}}(t)$.}
\label{fig:continuum_fracture}
\end{figure}

\subsubsection{Balance of Momentum}
In our fracture model, the balance of momentum equation can be simplified drastically compared to the full Navier-Stokes equations for viscous and compressible fluids. Studies on flow processes in hydraulically transmissive fractures with low contact areas motivate the simplification to a pressure-driven Poiseuille-type flow between two parallel plates \cite{witherspoon1980,vinci2014} under creeping flow conditions. Assuming constant or quasi-static fracture aperture, the geometrical characteristics of high-aspect-ratio fractures lead to predominant flow within the fracture plane. Integration of the velocity profile of the respective Poiseuille-type (assuming no-slip boundary conditions at the fracture surfaces) yields the relative fluid velocity 
\begin{equation}
\label{eq:hda_fluidflow}
\hat{\mathbf{w}}_\mathfrak{f} = -\frac{\aperture^2(\coordinateVec,t)}{12\,\fluidEffectiveDynViscosity} \, \hat{\text{grad}} \, \hat{p}
= -\frac{k^\mathfrak{s}_{Fr}(\coordinateVec,t)}{\fluidEffectiveDynViscosity} \, \hat{\text{grad}} \, \hat{p},
\end{equation}
where $\fluidEffectiveDynViscosity$ denotes the fluid’s effective dynamic viscosity and $k^\mathfrak{s}_{Fr}(\coordinateVec,t) := \aperture^2(\coordinateVec,t)/12$ the space and time resolved effective fracture permeability \footnote{Geometrical characteristics such as the fracture surface roughness can be governed by adaption of the proportionality factor $1/12$ \cite{renshaw1995}.}. Variables and mathematical operations defined by means of the fracture domain $\fractureDomain$ are denoted with $\hat{\msquare}$ to avoid confusion with quantities and higher dimensional operations defined in the poro-elastic domain $\poroElasticDomain$.

\subsubsection{Balance of Mass}
Mass conservation has to take into account volumetric changes and fluid compressibility $\beta^\mathfrak{f}$. For a given fluid compressibility $\beta^\mathfrak{f} = 1/K^\mathfrak{f}$ with respect to the fluid's bulk modulus $K^\mathfrak{f}$, the local mass balance reads

\begin{equation}
\label{eq:hda_balance_of_mass}
\pderiv{(\rho^{\mathfrak{f}R}\,\aperture)}{t} + \hat{\text{div}}
\left(\, \hat{\mathbf{w}}_\mathfrak{f} \, \rho^{\mathfrak{f}R}\, \aperture\right) = \text{w}_\mathfrak{f}^N,
\end{equation}
where $\text{w}_\mathfrak{f}^N$ is the leak-off triggered seepage velocity of the porous domain normal to the fracture domain.

\subsubsection{Governing Equations}
The governing equation for the flow in deformable fractures is obtained by combining \eq{\ref{eq:hda_fluidflow}} with the balance of mass in \eq{\ref{eq:hda_balance_of_mass}}. To close the set of governing equations, we introduce a linear equation of state for the fluid pressure 
$p \propto \rho^{\mathfrak{f}R}$ 
\begin{equation*}
p = K^\mathfrak{f} \,
\left[ 
\cfrac{\rho^{\mathfrak{f}R}}{\rho^{\mathfrak{f}R}_0} - 1
\right]
\end{equation*}
where $\rho^{\mathfrak{f}R}_0$ is the effective density at $t=0$. This results in the governing equation
\newcommand*{\termI}{\pderiv{\hat{p}}{t}}
\newcommand*{\termII}{\frac{\aperture^2}{12 \,\fluidEffectiveDynViscosity} \, \gradientLowDim \,\hat{p} \cdot \gradientLowDim \, \hat{p}}
\newcommand*{\termIII}{\frac{\aperture }{12\, \fluidEffectiveDynViscosity\, \beta^{\mathfrak{f}}} \gradientLowDim \, \aperture \cdot \gradientLowDim \, \hat{p}}
\newcommand*{\termIV}{\frac{1}{12\,\, \fluidEffectiveDynViscosity \beta^{\mathfrak{f}}} \divergenceLowDim \left( \aperture^2 \, \gradientLowDim \, \hat{p}\right)}
\newcommand*{\termV}{\frac{1}{\aperture\, \beta^{\mathfrak{f}}}\pderiv{\aperture}{t}}
\newcommand*{\termVI}{\text{w}_\mathfrak{f}^N/(\aperture\, \beta^\mathfrak{f})}
\begin{equation}
\label{eq:governing_hybrid}
  \def\vph{ \vphantom{\termI \termII \termIII \termIV \termV \termVI } }
\begin{aligned}
    { \underbrace{ \vph \termI   }_{ \I   } }
  - { \underbrace{ \vph \termII  }_{ \II  } }
  - { \underbrace{ \vph \termIII }_{ \III } }
  &
  \\
  - { \underbrace{ \vph \termIV  }_{ \IV  } }
  + { \underbrace{ \vph \termV   }_{ \V   } }
  &= 
    { \underbrace{ \vph \termVI  }_{ \VI  } }.
\end{aligned}
\end{equation}
Note that \eq{\ref{eq:governing_hybrid}} establishes a lower-dimensional non-linear partial differential equation for the pressure $\hat{p}$ i the fracture, i.e., for a scalar unknown function. 
It consists of a transient term $\I$, a quadratic term $\II$, a convection term $\III$, a diffusion term $\IV$, a volumetric coupling term $\V$ and a leak-off term $\VI$. The leak-off term $\hat{q}_{lk}=\text{w}_\mathfrak{f}^N/(\aperture\, \beta^\mathfrak{f})$ is related to the leak-off triggered seepage velocity $\text{w}_\mathfrak{f}^N$.

The authors of \cite{vinci2014} have shown via a dimensional analysis, that the terms $\II$ and $\III$ can be neglected in the regime of high aspect-ratio fractures due to their minor contribution to the overall solution. Thus, the final form of the fracture flow equation reads
\begin{equation}
\label{eq:governing_hybrid_simplified}
  \termI - \termIV + \termV = \termVI \; \mbox{ in } \fractureDomain.
\end{equation}
The reduced form introduced by this governing equation still consists of the pressure diffusion term $\II$ and the volumetric coupling term $\V$ which require an implicit, non-trivial coupling to the deformation state of the biphasic material.

\subsection{Partial Differential Equations in the Poro-Elastic Domain $\poroElasticDomain$}
Our fracture equation derived above calculates the pressure evolution for a given deformation and seepage velocity of the porous matrix. In this section, we present the respective biphasic poro-elastic formulation \cite{biot1941,Renner2015,wang-2000} for the porous medium.  
As the main emphasis of this work is on the hybrid-dimensional fracture flow formulation and its coupling to the porous medium, we introduce and explain the governing equations without any explicit derivation of the balance equations. A more detailed derivation of the poro-elastic formulation can be found in \cite{Schmidt2019}. 

\subsubsection{Governing Equations}
The governing equations of the poro-elastic mixture consisting of a solid and a pore-fluid constituent that are studied under quasi-static conditions. The unknown functions are the pore-fluid pressure $p$ and the solid displacement $\mathbf{u}_s$ given by the following equations:
\renewcommand{\arraystretch}{1.9}
\begin{equation}
\begin{gathered}
\label{eq:poro_governing}
\left.
\begin{array}{rcl}
-\text{div} (\boldsymbol{\sigma^\mathfrak{s}_E} - \alpha\,p\,\mathbf{I})  &=&  \rho\,\mathbf{b},\\
\cfrac{1}{M}\, \cfrac{\partial p}{\partial t} - \cfrac{k^\mathfrak{f}}{\gamma_0^{\mathfrak{f}R}} \,\text{div}\,\text{grad}\, p  &=&  -\alpha\,
\text{div} \, \cfrac{\partial \mathbf{u}_s}{\partial t},
\end{array} \qquad\right\}\; \mbox{ in } \; \poroElasticDomain, \\
\begin{array}{ll} 
\underline{\text{with:}} \qquad& \boldsymbol{\sigma}^\mathfrak{s}_E = 3\,K\,\text{vol}(\boldsymbol{\varepsilon}_\mathfrak{s}) +
2\,G\,\text{dev}(\boldsymbol{\varepsilon}_\mathfrak{s}) + (1-\alpha)\,p\,\mathbf{I}.
\end{array}
\end{gathered}
\end{equation}
\renewcommand{\arraystretch}{1.0}
The parameters are the Biot parameter $\alpha = 1 - K/K^\mathfrak{s}$, where $K$ is the dry bulk modulus of the solid skeleton, $K^\mathfrak{s}$ the bulk modulus of the compressible grains forming the skeleton and $\mathbf{b}$ are the body forces. The (local) storativity or inverse storage capacity is defined by $1/M = \phi_0 / K^\mathfrak{f} + (\alpha - \phi_0)/K^\mathfrak{s}$ with the the bulk modulus $K^\mathfrak{f}$ of the pore fluid and the porosity in the initial configuration $\phi_0 = \phi(t_0)$ of the mixture relating the partial and effective densities of the constituents. $k^\mathfrak{f}$ is the Darcy permeability or hydraulic conductivity, $\gamma_0^{\mathfrak{f}R}$ the effective weight at $t=t_0$, $\boldsymbol{\varepsilon}_\mathfrak{s}$ the elastic strain (split into a volumetric and a deviatoric part) and $G$ the dry shear modulus of the skeleton \cite{wang-2000,steeb-2019a}. 

Compressibility of the porous material is modeled following the assumption of linear poroelasticity \cite{biot1941}. Fluid-flow processes and solid deformations within the fracture and porous domain are now governed by \eq{\ref{eq:governing_hybrid_simplified}} and \eq{\ref{eq:poro_governing}}.

\subsection{Coupling Conditions}
\label{sec:equations:couplingConditions}

\begin{figure}[h]
\centering
\input{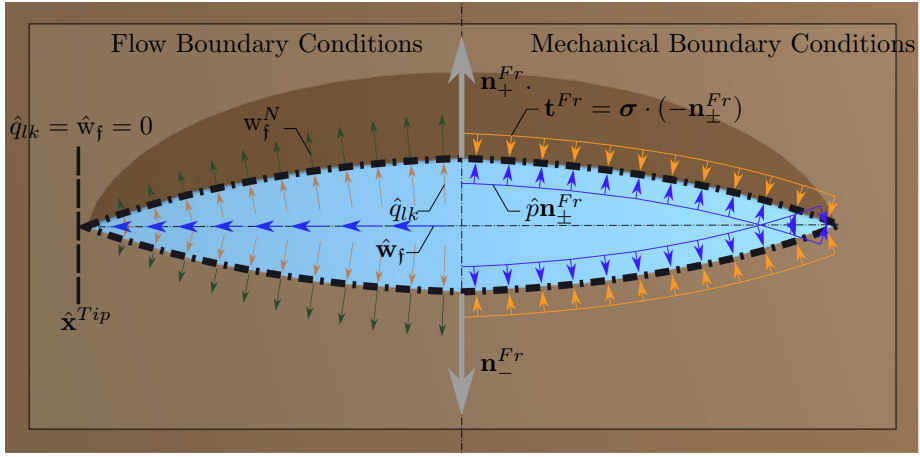}
\caption{Representation of flow and mechanical coupling conditions along the fracture surface $\fractureSurface$. Fracture normal vectors $\mathbf{n}^Fr_\pm$ are introduced and can vary along the fracture surface dependent on the evaluation position $\hat{\coordinateVec}$. Flow coupling conditions are characterized by the outflow $\hat{q}_{lk}=\termVI$ and normal seepage velocity $\text{w}_\mathfrak{f}^N$ in combination with a no-flow requirement at the fracture tips $\fractureTipLowDim$. Mechanical coupling conditions are described with respect to the fracture surface traction $\mathbf{t}^\text{Fr}$ and the acting fluid pressure $\hat{p}$.}
\label{fig:fracture_boundary}
\end{figure}

For a consistent coupling of the fracture and the poro-elastic domain, the system needs to be closed by coupling conditions along the fracture surface. The respective equilibrium conditions along the fracture surface $\fractureSurface$ as displayed in Fig.~\ref{fig:fracture_boundary} extend \eq{\ref{eq:governing_hybrid_simplified}} and \eq{\ref{eq:poro_governing}} to a well-defined system of equations for the whole simulation domain comprising fractures and surrounding porous matrix. 
The equilibrium state of the fluid flow is reached, once equivalent exchange between the fracture zone and the poro-elastic matrix becomes apparent and no-flow boundary conditions at the fracture tips $\fractureTipLowDim$ are met. Flow exchange between both domains is governed by leak-off flow $\hat{q}_{lk}$ within the fracture domain and the normal seepage velocity $\text{w}_\mathfrak{f}^N$ within the poro-elastic matrix. In this framework, considering a biphasic porous material, the normal seepage velocity is constructed based on the residual state obtained for a prescribed fluid pressure along the fracture surface matching the fluid pressure state of the fracture domain. Under the consideration of Green's theorem \cite{widlund2004}, the reconstruction is given by inserting  the obtained solution fields into
\begin{equation}
    \text{w}_\mathfrak{f}^N(\coordinateVec) = \cfrac{\dot{p}}{M} - \cfrac{k^\mathfrak{f}}{\gamma_0^{\mathfrak{f}R}} \,\text{div}\,\text{grad}\, p  + \alpha\,
\text{div} \,\dot{\mathbf{u}}_s \quad \mbox{at} \quad \fractureSurface.
    \label{eq:normal_seepage}
\end{equation}
The equilibrium conditions of the mechanical state are met once surface traction $\mathbf{t}^\text{Fr} = -\boldsymbol{\sigma} \cdot \mathbf{n}^\text{Fr}_\pm$, defined by the total stresses $\boldsymbol{\sigma}$, and the fluid pressure $\hat{p}$ acting normal to the fracture surface are balanced. Despite this definition of the coupled system's equilibrium state via surface terms, the physical nature of the coupling is volumetric. This is due to the fluid pressure $\hat{p}$ being relevant for both, the fluid and mechanical boundary conditions. It is defined by the evolution equation \eqref{eq:governing_hybrid_simplified} consisting of the volumetric coupling term $\V$. In conclusion, the necessary boundary conditions along the coupling interface are
\begin{subequations} 
  \label{eq:equations:coupling-conditions}
\begin{alignat}{3}
\label{eq:equations:coupling-condition-displacement}
\hat{q}_{lk} &= \text{w}_\mathfrak{f}^N/(\aperture\beta^\mathfrak{f}) \quad &\text{on} \quad \fractureSurface  &\qquad \text{and} \qquad 
\hat{\text{w}}_\mathfrak{s} = \hat{\text{w}}_\mathfrak{f}=0 \quad &\text{at} \quad \fractureTipLowDim,\\
\label{eq:equations:coupling-condition-pressure}
\mathbf{t}^\text{Fr} &= -\hat{p} \, \mathbf{n}^\text{Fr}_\pm \quad &\text{on} \quad \fractureSurface& \qquad \text{and} \qquad \hat{p} = p \quad &\text{on} \quad \fractureSurface.
\end{alignat}
\end{subequations} 

\subsection{Treatment of the Poro-Elastic Response on Different Time Scales}
Investigations of systems involving (crystalline) rock characterized by a low permeability, perturbations in the low frequency range ($\ll 100$ Hz), under undrained conditions, and with a low viscous pore fluid motivate the application of Gassmann's effective low-frequency result   \cite{gassmann1951,mavko2009}. Considering pore pressure effects by effective parameters in a single-phase, solid formulation introduces two major numerical advantages since a) the number of degrees of freedom considered in the rock domain is reduced and b) numerical instabilities due to distinct diffusion times of the rock and the fracture domain can be avoided. Still, investigation periods must be considerably smaller than the characteristic diffusion time of the rock matrix to ensure the negligible contribution of leak-off effects. The effective parameters based on Gassmann's effective low-frequency result are determined by
\begin{equation}
    \label{eq:gassmann}
    \begin{aligned}
    K_{\mathrm{eff}}&=\frac{\phi_0 \left(\frac{1}{K^\mathfrak{s}}-\frac{1}{K^\mathfrak{f}}\right) + \frac{1}{K^\mathfrak{s}} -\frac{1}{K}}{\frac{\phi_0}{K}\left(\frac{1}{K^\mathfrak{s}}-\frac{1}{K^\mathfrak{f}}\right) + \frac{1}{K^\mathfrak{s}}\left( \frac{1}{K^\mathfrak{s}} - \frac{1}{K} \right)}, \\[+3mm]
    G_{\mathrm{eff}} &= G.
    \end{aligned}
\end{equation}
Here, $K_{\mathrm{eff}}$ is the effective bulk and $G_{\mathrm{eff}}$ the effective shear modulus. The poro-elastic effects are governed by the effective Gassman modulus $K_{\mathrm{eff}}$.
In this work, we assume isotropic linear elasticity of the surrounding bulk material (with effective elastic parameters $K_{\mathrm{eff}} $ and $G_{\mathrm{eff}} $) to avoid numerical instabilities in the pore pressure solution once the introduced coupling conditions eq.~\eqref{eq:equations:coupling-conditions} are fulfilled.

\section{Partitioned coupling}
\label{sec:couplig}
The coupling of the poro-elastic medium and the fracture flow is realized using partitioned coupling methods instead of a monolithic solver, i.e., we solve the equations for the fracture domain $\fractureDomain$ \eqref{eq:governing_hybrid_simplified} and poro-elastic domain $\poroElasticDomain$ \eqref{eq:poro_governing} separately. The coupling conditions \eq{\ref{eq:equations:coupling-condition-displacement}} and \eq{\ref{eq:equations:coupling-condition-pressure}} are enforced via suitable boundary conditions for the subdomains, see Fig.~\ref{fig:fracture_boundary}, and iterative exchange of the respective boundary values within a time step. 

We use a fracture solver $F$ that maps a seepage velocity $\text{w}_\mathfrak{f}^N$ and an aperture $\aperture$ to a pressure $p$ by executing a discrete time step for \eq{\ref{eq:governing_hybrid_simplified}}. The corresponding porous medium solver $S$ maps the pressure $p$ at the fractures back to a seepage velocity $\text{w}_\mathfrak{f}^N$ and an aperture $\aperture$ by executing a time step for \eq{\ref{eq:poro_governing}} while deriving $\text{w}_\mathfrak{f}^N$ and $\aperture$ via \eq{\ref{eq:normal_seepage}} and the difference between displacements $\displacement^{\pm}$ at both sides of the fracture. This can be considered equivalent to the typical approach in classical fluid-structure interactions (FSI) with elastic solid structures as Dirichlet-Neumann domain decomposition approach \cite{Deparis2006,Farhat2004,Monge2018}. Note that there are two major differences compared to classical FSI: (i) the fluid domain is modeled with a lower-dimensional simplified equation and, thus, the transferred aperture and seepage velocities are not boundary conditions in the strict sense; (ii) the fluid exchange between porous matrix and fractures and the compressibility of the porous structure and the fluid might be a factor that makes our problem less prone to instabilities than FSI, in particular with incompressible fluids. A potential third point are the relatively large fracture aperture changes that appear in the first time step for most applied boundary conditions. This makes the coupled problem hard to solve in the very beginning which is also the case for the studied test cases in this work.

By iteratively solving one of the fixed-point equations
\begin{subequations}
\label{eq:fixed_point}
\begin{eqnarray}
S \circ F( \aperture, \text{w}_\mathfrak{s}^N) & = & (\aperture, \text{w}_\mathfrak{s}^N) \; \mbox{ or } \label{eq:fixed_point_Jacobi} 
\\[+3mm]
\left( \begin{array}{c} F( \aperture, \text{w}_\mathfrak{s}^N) \\ S ( p ) \end{array} \right) & = & 
\left( \begin{array}{c} p \\ ( \aperture, \text{w}_\mathfrak{s}^N) \end{array} \right), \label{eq:fixed_point_GS}
\end{eqnarray}
\end{subequations}
we, thus, ensure fulfilment of all equations and coupling conditions at the new time step. In this formulation, input and output of $F$ and $S$ refer to the end point of the respective time step.

One of the main advantages of the approach presented here is that it allows for  \enquote{black box} coupling, i.e., we can reuse existing solvers that have been tailored for the problems in the subdomains. If we can provide an efficient iterative scheme to solve \eq{\ref{eq:fixed_point}}, we do not have to develop a dedicated solver for the overall highly ill-conditioned system of equations that arises from the monolithic approach, either. Note that, despite of the equilibrium state definition for the coupled system via the fracture surface, the nature of the coupling from the point of view of the fracture domain $\fractureDomain$ is volumetric. This is due to the involvement of the fluid pressure $\hat{p}$ in both, the fluid and mechanical boundary conditions, which is defined by evolution equation \eqref{eq:governing_hybrid} incorporating the volumetric coupling term $\V$. 

\subsection{Iterative Coupling}
We present different options to solve \eq{\ref{eq:fixed_point}} within each time step. These options have been presented before in~\cite{Scheufele2017,preCICE,Degroote2008,Degroote2010} and evaluated for classical FSI problems. In this work, we analyze their potential for the lower-dimensional fracture flow problem. The cheapest version in terms of cost per time step, but in general known to generate unstable time stepping is the \emph{explicit coupling} were each solver is executed only once per time step, i.e., 
\begin{subequations}
  \begin{align}
  (\aperture^{(n+1)}, \text{w}_\mathfrak{s}^{N,(n+1)}) &= S \circ \underbrace{F( \aperture^{(n)}, \text{w}_\mathfrak{s}^{N,(n)})}_{\mbox{$=: p^{(n+1)}$}}   \; \mbox{ or } 
  \\
  \left( \begin{array}{c} p^{(n+1)} \\ ( \aperture^{(n+1)}, \text{w}_\mathfrak{s}^{N,(n+1)}) \end{array} \right) &= \left( \begin{array}{c} F( \aperture^{(n)}, \text{w}_\mathfrak{s}^{N,(n)}) \\ S ( p^{(n)} ) \end{array} \right),
  \label{eq:fixed_point_explicit}
  \end{align}
\end{subequations}
where the superscript $(n)$ denotes the discrete solution at time $t_n$. Note that the first option in \eq{\ref{eq:fixed_point_explicit}} implies a sequential one-after-the-other execution of the two solvers $F$ and $S$, where $S$ already uses the ``new'' pressure $p^{(n+1)}$ as an input, whereas, in the second option, both solvers can be executed simultaneously. Explicit coupling cannot capture the strong physical interaction between the fracture and porous matrix and is not considered further.

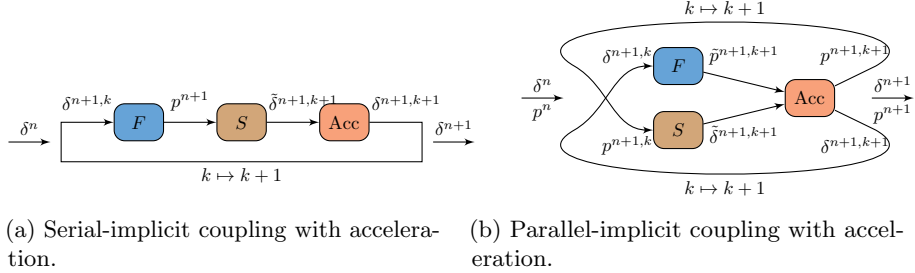
\begin{figure}
  \begin{subfigure}[b]{0.47\textwidth}
    \centering
      \begin{tikzpicture}[scale=0.7, transform shape,  node distance = 0.75cm, auto]
  \node[flowSolver] (flowSolver) {$\flowSolverLabel$};
  \node[solidSolver, right=1.0 cm of flowSolver] (solidSolver) {$\solidSolverLabel$};
  
  \node[accelerator, right=1.0 cm of solidSolver] (accelerator) {\acceleratorLabel{}};
  
  \path[line] (flowSolver) -- node[above, yshift=1mm] {$\pressureNewExpl$} (solidSolver);
  \path[line] (solidSolver) -- node[above, xshift=2mm,yshift=1mm] {$\apertureNewFirstImpl$} (accelerator);
  
  \node[coord, right=1cm of solidSolver] (sR) {};
  \node[coord, below of=sR] (sBR) {};
  \node[coord, left= 1cm of flowSolver] (fL) {};
  \node[coord, below of=fL] (fBL) {};
  
  \node[coord, right=1cm of accelerator] (aR) {};
  \node[coord, below of=aR] (aBR) {};
  
  \path[line] (accelerator) -- node[above, xshift=2mm,yshift=1mm] {$\apertureNewImpl$} (aR) -- (aBR) -- node[below] {$\iterIndexExpl \mapsto \iterIndexExpl+1$} (fBL) -- (fL) -- node[above, yshift=1mm] {$\apertureOldImpl$} (flowSolver);
  
  \node[coord,xshift=-1.0cm] (incomingArrowLeft) at ($(fL)!0.5!(fBL)$) {};
  \node[coord,xshift=-0.2cm] (incomingArrowRight) at ($(fL)!0.5!(fBL)$) {};
  
  \path[line] (incomingArrowLeft) -- node[above] {$\apertureOldExpl$} (incomingArrowRight);
  
  \node[coord,xshift=0.21cm] (outgoingArrowLeft) at ($(aR)!0.5!(aBR)$) {};
  \node[coord,xshift=1.0cm] (outgoingArrowRight) at ($(aR)!0.5!(aBR)$) {};
  
  \path[line] (outgoingArrowLeft) -- node[above] {$\apertureNewExpl$} (outgoingArrowRight);
  
  \draw[draw=none] (-1,-1.5) rectangle (1,0.75);
  \end{tikzpicture}
    \subcaption{Serial-implicit coupling with acceleration.}
    \label{fig:coupling:serial-implicit}
  \end{subfigure}
  \quad
  \begin{subfigure}[b]{0.47\textwidth}
    \centering
    \begin{tikzpicture}[scale=0.7, transform shape, node distance = 0.75cm, auto]
\node[flowSolver] (flowSolver) {$\flowSolverLabel$};
\node[solidSolver, below=0.5 cm of flowSolver] (solidSolver) {$\solidSolverLabel$} ;

\node[accelerator, xshift=2.5cm] (accelerator) at ($(flowSolver)!0.5!(solidSolver)$) {\acceleratorLabel{}};

\node[coord, right=1cm of accelerator] (aR) {};

\node[coord, right of=solidSolver, xshift=3.0cm] (aBR) {};
\node[coord, right of=flowSolver, xshift=3.0cm] (aTR) {};

\node[coord, left=1.5cm of solidSolver] (sL) {};
\node[coord, left= 1.5cm of flowSolver] (fL) {};

\node[coord, right=1cm of solidSolver] (sR) {};
\node[coord, right= 1cm of flowSolver] (fR) {};

\node[coord] (center) at ($(flowSolver)!0.5!(solidSolver)$) {};

\path[line] (flowSolver) -- node[above,yshift=2mm] {$\pressureNewFirstImpl$} (accelerator);

\path[line] (solidSolver) -- node[below,yshift=-1mm] {$\apertureNewFirstImpl$} (accelerator);

\path[line] (accelerator)  -- node[above,yshift=1.0mm] {$\pressureNewImpl$} (aTR)  to[out=30,in=150] node[above]   {$\iterIndexExpl\mapsto \iterIndexExpl+1$} (fL) to[out=-30,in=180]  node[below,yshift=-4.5mm,xshift=3mm]  {$\pressureOldImpl$} (solidSolver);

\path[line] (accelerator)  -- node[below,yshift=-2.0mm] {$\apertureNewImpl$} (aBR)  to[out=-30,in=-150] node[below]   {$\iterIndexExpl\mapsto \iterIndexExpl+1$} (sL) to[out=30,in=180]  node[above,yshift=4mm,xshift=3mm]  {$\apertureOldImpl$} (flowSolver);

\node[coord,xshift=-1.0cm] (incomingArrowLeft) at ($(fL)!0.5!(sL)$) {};
\node[coord,xshift=-0.2cm] (incomingArrowRight) at ($(fL)!0.5!(sL)$) {};

\path[line] (incomingArrowLeft) -- node[above] {$\apertureOldExpl$} node[below] {$\pressureOldExpl$} (incomingArrowRight);

\node[coord,xshift=-0.3cm] (outgoingArrowLeft) at ($(aR)$) {};
\node[coord,xshift=0.5cm] (outgoingArrowRight) at ($(aR)$) {};

\path[line] (outgoingArrowLeft) -- node[above] {$\apertureNewExpl$} node[below] {$\pressureNewExpl$} (outgoingArrowRight);

\end{tikzpicture}
    \subcaption{Parallel-implicit coupling with acceleration.}
    \label{fig:coupling:parallel-implicit}
  \end{subfigure}
  \caption{Sketch of the used coupling schemes. For the sake of readability, we display the case without the normal seepage velocity $\text{w}_\mathfrak{s}^{N}$.}
  \label{fig:coupling:implicitCouplingSchemes}
\end{figure}

\emph{Implicit coupling} can be achieved via fixed point iterations
\begin{subequations}
\begin{align}
(\aperture^{(n+1),i+1}, \text{w}_\mathfrak{s}^{N,(n+1),i+1}) 
&= S \circ \underbrace{F( \aperture^{(n+1),i}, \text{w}_\mathfrak{s}^{N,(n+1),i})}_{\mbox{$=: p^{(n+1),i+1}$}} \; \mbox{ or }
\\
\left( 
\begin{array}{c} 
  p^{(n+1),i+1} \\
  ( \aperture^{(n+1),i+1}, \text{w}_\mathfrak{s}^{N,(n+1),i+1}) 
\end{array} 
\right) 
&= 
\left( 
\begin{array}{c} 
  F( \aperture^{(n+1),i}, \text{w}_\mathfrak{s}^{N,(n+1),i}) \\ 
  S ( p^{(n+1),i} )
\end{array} 
\right).
\label{eq:fixed_point_implicit}
\end{align}
\end{subequations}
The first equation refers to a \emph{serial-implicit}, cf.~\figref{fig:coupling:serial-implicit}, and the second equation to a \emph{parallel-implicit} coupling, cf.~\figref{fig:coupling:parallel-implicit}.
In the figures, an additional element, the so-called acceleration scheme, is shown. This numerical component generates an improved next iterate based on the output of the respective fixed-point iteration either by under-relaxation or interface quasi-Newton (IQN) methods. To simplify the notation in the following, we use the general formulation
\begin{equation}
  \fixedPointProblemFunc{\surfaceUnknown} = \surfaceUnknown
  \label{eq:coupling:fixed-point-problem}
\end{equation}
with a vector of unknowns $\surfaceUnknown \in \setOfRealNumbers^\surfaceProblemDim$ at the fractures and the fixed-point operator $\fixedPointProblem: \setOfRealNumbers^\surfaceProblemDim \mapsto \setOfRealNumbers^\surfaceProblemDim$ for both variants of fixed-point equations in eq.~\ref{eq:fixed_point}. 

We write the general accelerated version of the fixed-point iterations in \eq{\ref{eq:fixed_point_implicit}} as
\[\surfaceUnknown^{i+1} = \mathcal{A}(\fixedPointProblemFunc{\surfaceUnknown^i})\,,\]
with the acceleration operator $\mathcal{A}$. A straight-forward option to realize the acceleration is to use underrelaxation,
\begin{equation}
\surfaceUnknown^{i+1} = \omega \fixedPointProblemFunc{\surfaceUnknown^i} + (1-\omega) \surfaceUnknown^i \,,\omega \in ]0;1]\,,
\label{eq:underrelaxation}
\end{equation}
e.g., based on a dynamic Aitken scheme \cite{Kuettler2008}. A more advanced approach is to reuse past iterates to establish quasi-Newton iterations \cite{Degroote2008,Degroote2010, Haelterman2009}\footnote{To avoid linear dependencies between information from previous iterations, modified Newton iterations starting from the result of the pure fixed-point iteration are used (for details, see 
\cite{Uekermann2016}).}:
\begin{equation*}
  \surfaceUnknown^{i+1} = \tilde{\surfaceUnknown}^{i} + \Delta \tilde{\surfaceUnknown}^{i} \;\;\text{with}\;\; \Delta \tilde{\surfaceUnknown}^{i} = J_{\tilde{R}}^{-1}(\Delta \tilde{\surfaceUnknown}^{i}) \tilde{R}(\Delta \tilde{\surfaceUnknown}^{i}),
\end{equation*}
with $\tilde{\surfaceUnknown}^{i}:=\fixedPointProblemFunc{\tilde{\surfaceUnknown}^{i}}$, the modified residual $\tilde{R}(\tilde{\surfaceUnknown}):= \tilde{\surfaceUnknown} - \fixedPointProblemFuncInv{\tilde{\surfaceUnknown}}$, and $J_{\tilde{R}}^{-1}(\tilde{\surfaceUnknown}^i)$ as an approximation of the inverse of the Jacobian of $\tilde{R}$. 

To compute $J_{\tilde{R}}^{-1}(\tilde{\surfaceUnknown}^i)$, we collect input-output information from past iterates of $\fixedPointProblem$ in tall and skinny matrices $V_i, W_i \in \mathbb{R}^{m \times i}$, $m \gg i$,
\begin{eqnarray*}
   V_i &=& \left[\tilde{R}(\tilde{\surfaceUnknown}^1) - \tilde{R}(\tilde{\surfaceUnknown}^0), \tilde{R}(\tilde{\surfaceUnknown}^2) - \tilde{R}(\tilde{\surfaceUnknown}^1), \ldots, \tilde{R}(\tilde{\surfaceUnknown}^i) - \tilde{R}(\surfaceUnknown^{i-1}) \right]\;, \\
   W_i &=& \left[\tilde{\surfaceUnknown}^1 - \tilde{\surfaceUnknown}^0, \tilde{\surfaceUnknown}^2 - \tilde{\surfaceUnknown}^1, \ldots, \tilde{\surfaceUnknown}^i - \tilde{\surfaceUnknown}^{i-1} \right]\;.
\end{eqnarray*}
The matrices $V_i$ and $W_i$ define the multi-secant equations for the inverse Jacobian 
\begin{equation}
\label{equ::secant}
J_{\tilde{R}}^{-1} (\tilde{\surfaceUnknown}^i) \; V_i = W_i\;.
\end{equation}
To get the classical interface quasi-Newton inverse least-squares (IQN-ILS) method \cite{Degroote2010,Degroote2008}, we close \eqref{equ::secant} by 
\[\|J^{-1}_{\tilde{R}}(\tilde{\surfaceUnknown}^i)\|_F \rightarrow \text{min} \,.\]
For time-dependent problem, we have to solve the coupling problem for every time step $n$. We can extent the notation of $\Vi$ and $\Wi$ to $\Vin$ and $\Win$ to emphasize the collection of differences in the current time step. In this case, the convergence of the quasi-Newton method can be improved by additionally consider information of previous time steps, i.e.\ to use $\Vi^{n-1}, \Vi^{n-2}\dots$ and $\Wi^{n-1}, \Wi^{n-2}\dots$. This is referred to as \emph{reuse} of time steps method. However, we cannot store information from an infinite amount of time due to memory restrictions and since information can be outdated. Thus, we define a reuse  parameter $\timeReuseParameter$ that defines for how long information is retained, i.e., we keep $\Vi^{n-1}, \Vi^{n-2},\dots,\Vi^{n-\timeReuseParameter}$ and $\Wi^{n-1}, \Wi^{n-2},\dots, \Wi^{n-\timeReuseParameter}$. If there are many coupling iterations per time step and, thus, the matrices $\Vin$ and $\Win$ are very large, this can lead to excessive memory requirements as well. To avoid storing the information of too many time steps, we also define the iteration reuse parameter $\iterationReuseParameter$ that limits how many data pairs over previous coupling iterations and time steps may be kept in total.

An alternative to IQN-ILS is based on the multi-vector (MV) approach \cite{Bogaers2014}. 
The multi-secant equation \eqref{equ::secant} is closed by 
\begin{equation*}
  \| J^{-1}_{\tilde{R}}(\tilde{\surfaceUnknown}^{i}) - J^{-1}_{\tilde{R}}(\tilde{\surfaceUnknown}^{\text{prev}}) \|_F \rightarrow \text{min} \, 
\end{equation*}
where $J^{-1}_{\tilde{R}}(\tilde{\surfaceUnknown}^{\text{prev}}$ is the approximation of the inverse Jacobian from the previous time step. The method is also referred to as interface quasi-Newton inverse multi-vector Jacobian (IQN-IMVJ) method \cite{Scheufele2017}. 
In contrast to the IQN-ILS method, we do not need to store information from previous time steps explicitly, since the information is kept by the explicit incorporation of $J^{-1}_{\tilde{R}}(\tilde{\surfaceUnknown}^{\text{prev}})$ in the minimization condition. However, this also leads to increased memory and runtime requirements for an increasing number of time steps, as the amount of information stored in $J^{-1}_{\tilde{R}}(\tilde{\surfaceUnknown}^{i}) $ and, thus, its size increases. Therefore, we use a IMVJ flavor with periodic restart to reduce runtime complexity and storage requirements. 

After number $\imvjRestart$ time steps, the method computes a singular value decomposition (SVD) of the approximated Jacobian and drops all information connected to singular values smaller than a user-specified threshold $\svdTrunctationThreshold$. We refer to this restart method as RS-SVD (restarted via singular value decomposition). For a detailed derivation and analysis of this restarted method, we refer to \cite{Scheufele2017}. Main advantages over the ILS method are the low number of parameters and that it showed less dependency on the choice of coupling parameters.

Both quasi-Newton methods may suffer from a lack of stability, if the columns in $\Vi$ become (nearly) linearly dependent. Therefore, additional stabilization of the numerical method is achieved by filtering to remove such nearly linearly dependent columns. In this work, we use a QR filter, also called QR2 filter, which constructs the QR decomposition $QR = V_i$ and drops columns that do not add sufficiently much new information to the problem based on a user-defined filter limit $\filterLimit$. A detailed description of this filtering technique and a comparison with other filter techniques can found in literature such as \cite{Haelterman2016,Scheufele2017}.

Implementation details of the partitioned coupling methods, including the efficient parallelization, data-mapping techniques etc., are out of the scope of this work. Instead, we refer to the reference paper of \preCICE{} \cite{preCICE} and the references therein.

\section{Numerical results}
\label{sec:results}
Computational efficiency and flexibility of the proposed partitioned coupling algorithm allow numerical investigations of non-trivial flow processes in deformable, arbitrarily fractured porous media in three dimensions for realistic initial aperture openings in the micrometer range. The capacity of the proposed method is shown throughout a number of numerical studies. First, the implementation of the partitioned scheme is verified by a  comparison to a reference solution (computed by a monolithic approach) for a simple boundary value problem consisting of a single embedded fracture. Afterwards, the efficiency and robustness of the method is explored  carrying out a study on the convergence behavior of the interface quasi-Newton schemes and their dependence on coupling parameters such as reuse $\timeReuseParameter$, restart $\imvjRestart$, and filter limit $\filterLimit$. The mesh dependency of the solution is investigated via parallel computations on meshes ranging from tens of thousand to several million degrees of freedom. The work is closed by demonstrating the potential of the approach to answer relevant questions in a broad range of fields including the inverse analysis of pumping operations and investigations related to nuclear waste disposal; two fields that clearly require computations with distinct boundary conditions and time scales.

The implementation of the proposed strategy uses the open-source libraries \preCICE{} and FEniCS to couple flow processes in the fracture domain $\fractureDomain_\text{s}$ governed by eq. (\ref{eq:governing_hybrid_simplified}) with responses of the poro-elastic domain $\poroElasticDomain_\text{s}$ defined by eqs. (\ref{eq:poro_governing}), resulting in a non-linear system. In both domains, the governing equations are solved by standard continuous Galerkin methods \cite{Schmidt2019}. 
\subsection{Verification of the partitioned implementation}
\label{sec:results_verification}
\begin{table*}[tb]
\caption{Parameters defining the validation boundary value problem.}
\centering
\begin{adjustbox}{max width=\textwidth}
\begin{tabular}{llllll}\hline
\rule[1.9ex]{0ex}{1ex}\bf{Quantity} & \bf{Value} & \bf{Unit} & \bf{Quantity} & \bf{Value} & \bf{Unit} \\[1.1ex]\hline
\textbf{\textit{Numerical parameters}} &&&&&  \\
solid depth $\text{d}_{\poroElasticDomain_\text{s}}$ & $10.0$ & [m] & solid width $\text{w}_{\poroElasticDomain_\text{s}}$ & $10.0$ &  [m] \\
solid height $\text{h}_{\poroElasticDomain_\text{s}}$ & $10.0$ & [m] &fracture radius $\text{r}_{\fractureDomain_\text{s}}$ & $1.25$ &  [m]\\
well diameter $\text{d}_{w}$ & $0.28$ & [m] &
solid vertices part. & $8.3\cdot 10^{4}$ &  [-] \\
fracture vertices part.  & $3.8 \cdot 10^{3}$ & [-] & solid vertices mono.  & $4.4 \cdot 10^{4}$ & [-] \\
fracture vertices mono.  & $1.0 \cdot 10^{3}$ & [-]  & time step size $\Delta t$ & $0.1$ & [sec] 
 \\
\textbf{\textit{Rock parameters}} &&&&&  \\
dry frame bulk modulus $K$ & $8.0$ & [GPa] & grain bulk modulus $K^\mathfrak{s}$ & $33.0$ & [GPa] \\
shear modulus $G$ & $15.0$ & [GPa] & initial porosity $\phi_0$ & $0.01$ & [-]  \\
intrinsic permeability $k^\mathfrak{s}$ & $1.1 \cdot 10^{-19}$  & [$\text{m}^2$] &
fluid compressibility $\beta^f$  & $0.45$  & [1/GPa] \\
effective bulk modulus $K_{eff}$ & $29.1$  & [GPa] &  effective shear modulus $G_{eff}$ & $15.4$  & [GPa] \\
\textbf{\textit{Fracture parameters}} &&&&&  \\
initial aperture $\delta_0$ & $50.0 $ & [$\mu$m] & effective fluid viscosity $\eta^{\mathfrak{f}R}$ & $0.001$ & [Pa$\cdot$s] \\
fluid compressibility $\beta^f$  & $0.45$  & [1/GPa] & injection pressure $\hat{p}_\text{i}^\text{s}$ & $200.0$ &  [kPa]\\
\textbf{\textit{Coupling Parameters}} &&&&&  \\
coupling & serial-implicit &  & quasi-Newton  & ILS &  \\
filter & QR2 & & initial relaxation $\omega$ & 0.001 &  [-]\\
rel. convergence & $10^{-6}$  & [-] &  & &  \\
\hline
\end{tabular}
\end{adjustbox}
\label{tab:validation}
\end{table*}
\begin{figure}[htb]
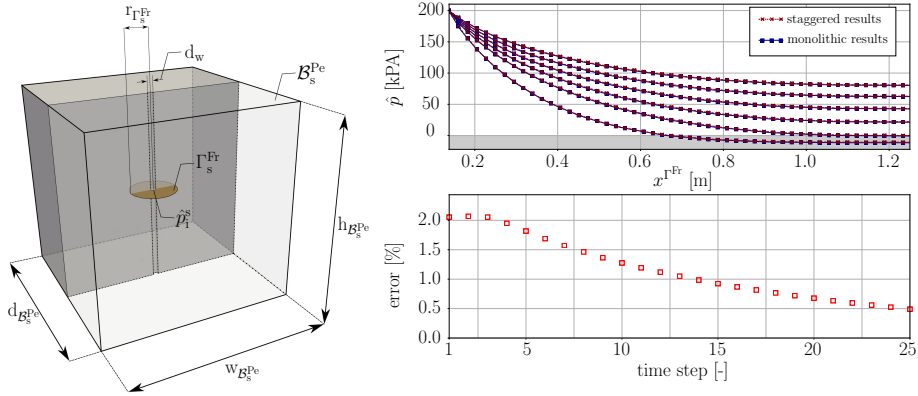

\begin{minipage}{0.4\textwidth}
\centering
    \resizebox{1.0\textwidth}{!}{\input{fig/numres/geometry_single_fracture.eps_tex}}
   \end{minipage}\hfill
   \begin{minipage}{0.6\textwidth}
\centering
    \resizebox{1.0\textwidth}{!}{\input{fig/numres/single_surface_results_without_disc_err.eps_tex}}
   \end{minipage}
     \caption{\textbf{Left:} Sketch of a single fracture embedded in a solid matrix used for comparisons of the monolithic and the partitioned scheme. \textbf{Top Right:} Pressure results obtained from both methods plotted over the fracture radius for time steps $t_i$ with $i = 1, 5, 10, 15, 20, 25$. \textbf{Bottom Right:} Relative error plots based on the deviation between solutions obtained from the monolithic and the partitioned scheme.}
     \label{Fig:res_comparison}
\end{figure}
Closed form solutions for non-linear flow processes in deformable fractures are not known to exist. Thus, the partitioned approach is verified by comparison to a monolithic scheme. 

Focusing on the hydro-mechanical interaction within the fracture domain, the surrounding bulk material is assumed to be linear-elastic where poro-elastic effects are transferred to the material parameters using Gassmann's effective low-frequency result, see eq. (\ref{eq:gassmann}). The parameters used in the investigated boundary value problem are introduced presented in Tab.~\ref{tab:validation} and the geometrical set up is given in \figref{Fig:res_comparison}. Throughout all numerical studies, Dirichlet deformation boundary conditions are applied to the poro-elastic domain $\poroElasticDomain$ by setting deformations on the outer surfaces in normal direction equal to zero. Within the fracture domain $\fractureDomain_\text{s}$, pressure Dirichlet conditions $\hat{p}_i^\text{s}$ are applied at the intersection of fracture and well. 
In \figref{Fig:res_comparison}, we compare the monolithic and partitioned solutions. We exploit the radial-symmetric characteristic of the boundary value problem and plot the pressure over the fracture radius at six different times. We observe very good agreement of the pressure for the monolithic and partitioned approach for all time steps. Additionally, we give the deviations between both strategies evaluated by a relative error
\begin{equation}
\label{eq:error}
\text{error}
=
\frac{\sum_{i=1}^{\text{N}_{\mathrm{c}}} \abs{\hat{p}_i^\mathrm{m} - \hat{p}_i^\mathrm{p}} }{\sum_{i=1}^{N_{\mathrm{c}}} \abs{\hat{p}_i^\mathrm{m}}}\cdot100  
\end{equation}
The pressure solution of both approaches has been interpolated to $N_c=50$ control points equally distributed over the fracture radius. The monolithic approach is based on quad meshes, while the partitioned approach is based on tetrahedral (porous matrix) and triangular (fracture) meshes. $3\,808$ degrees of freedom (DoF) were used in the fracture and $83\,109$ DoF in the solid domain for the partitioned approach and $44\,418$ DoF for the monolithic approach, where the fracture flow domain consists of $952$ DoF. Due to this, the elements and vertices of the monolithic and the partitioned coupling strategy do not match and a comparison of the obtained solutions can not be expected to result in perfect agreement.   Nevertheless, the results given by \figref{Fig:res_comparison} show reasonably small errors below $2.5\,\%$ and decrease over time. 

From a physical perspective, the obtained results are sound and the expected response in form of reverse water level fluctuations can be observed at an early stage, where negative pressure values are induced by non-local volumetric deformations of the fracture. Summarizing, the verification indicates the convergence of both coupling approaches towards the same solution.

\subsubsection{Investigation of interface quasi-Newton schemes}
\label{sec:results_parallel}
\begin{table*}[tb]
  \caption{Coupling parameters used in the study of quasi-Newton methods.}
  \centering
  \begin{adjustbox}{max width=\textwidth}
    \begin{tabular}{llllll}\hline
      \rule[1.9ex]{0ex}{1ex}\bf{Quantity} & \bf{Value} & \bf{Unit} & \bf{Quantity} & \bf{Value} & \bf{Unit} \\[1.1ex]\hline
      \textbf{\textit{Coupling Parameters}} &&&&&  \\
      coupling & $\lbrace\text{serial-implicit},\text{parallel-implicit}\rbrace$ &  & quasi-Newton  & $\lbrace\text{ILS},\text{IMVJ}\rbrace$ &  \\
      filter & QR2 & & initial relaxation $\omega$ & 0.1 &  [-]\\
      filter limit $\filterLimit$& $\lbrace 10^{-4}, 10^{-3}, 10^{-2}, 10^{-1}\rbrace$ & & relative convergence & $10^{-3}$  & [-]\\
      \textbf{\textit{ILS parameters}} &&&&&  \\
      Reuse parameter $\timeReuseParameter$ & $\lbrace 0, 8, 16, \infty \rbrace$ & [-] & & & \\      
      \textbf{\textit{IMVJ parameters}} &&&&&  \\
      SVD threshold $\svdTrunctationThreshold$ & $\lbrace 10^{-4}, 10^{-3}, 10^{-2}, 10^{-1}\rbrace$ & [-] & Restart parameter $\imvjRestart$ & 8 &  [-]\\
      \hline
    \end{tabular}
  \end{adjustbox}
  \label{tab:quasi-Newton:parameters}
\end{table*}

To verify the suitability of the partitioned coupling approach, we run a parameter study with the same settings as in Tab.~\ref{tab:validation} for the physical setup and coupling parameters as given in Tab.~\ref{tab:quasi-Newton:parameters}. All simulations are run in serial-implicit mode, i.e., we iterate in every time step until we fulfill the first fixed-point equation in eqs.~\ref{eq:fixed_point}. The initial relaxation is $\relaxationParameter=0.1$ and we use the QR2 filter. In order to keep the simulation feasible, we use a somewhat coarse mesh with $3\,292$ degrees of freedom in the fracture domain and $68\,853$ degrees of freedom in the porous-medium domain. If the reuse parameter is set to $\timeReuseParameter=\infty$, we allow the ILS method to keep information from all time steps. In this case, information from the current or previous time step is only dropped due to the QR2 filter.
\begin{figure}
  \centering
  \begin{subfigure}[c]{0.47\textwidth}
    \centering
    \includegraphics[keepaspectratio,width=1.1\linewidth]{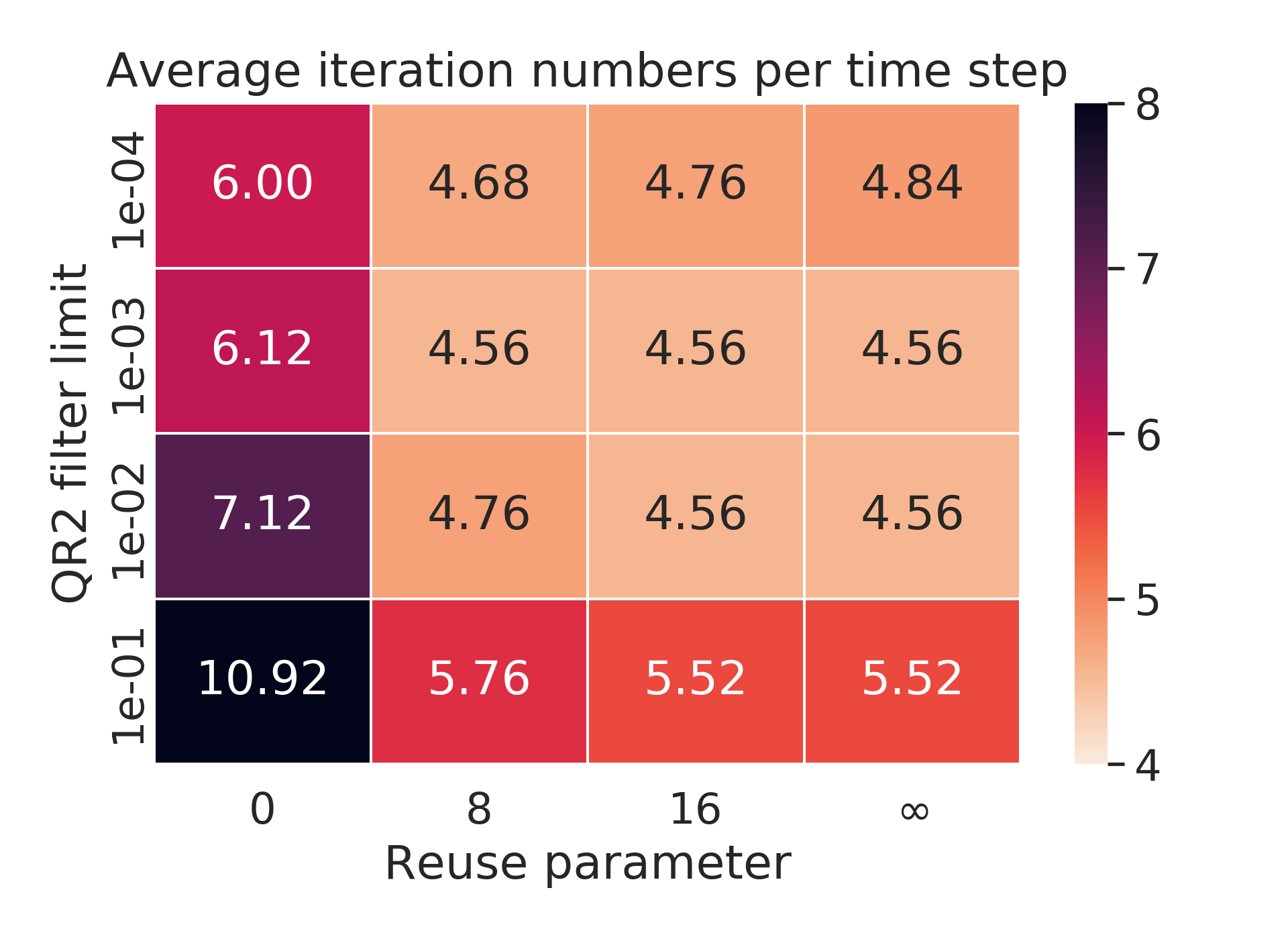}
    \caption{Using $\displacement^\pm$}
  \end{subfigure}
  \begin{subfigure}[c]{0.47\textwidth}
    \centering
    \includegraphics[keepaspectratio,width=1.1\linewidth]{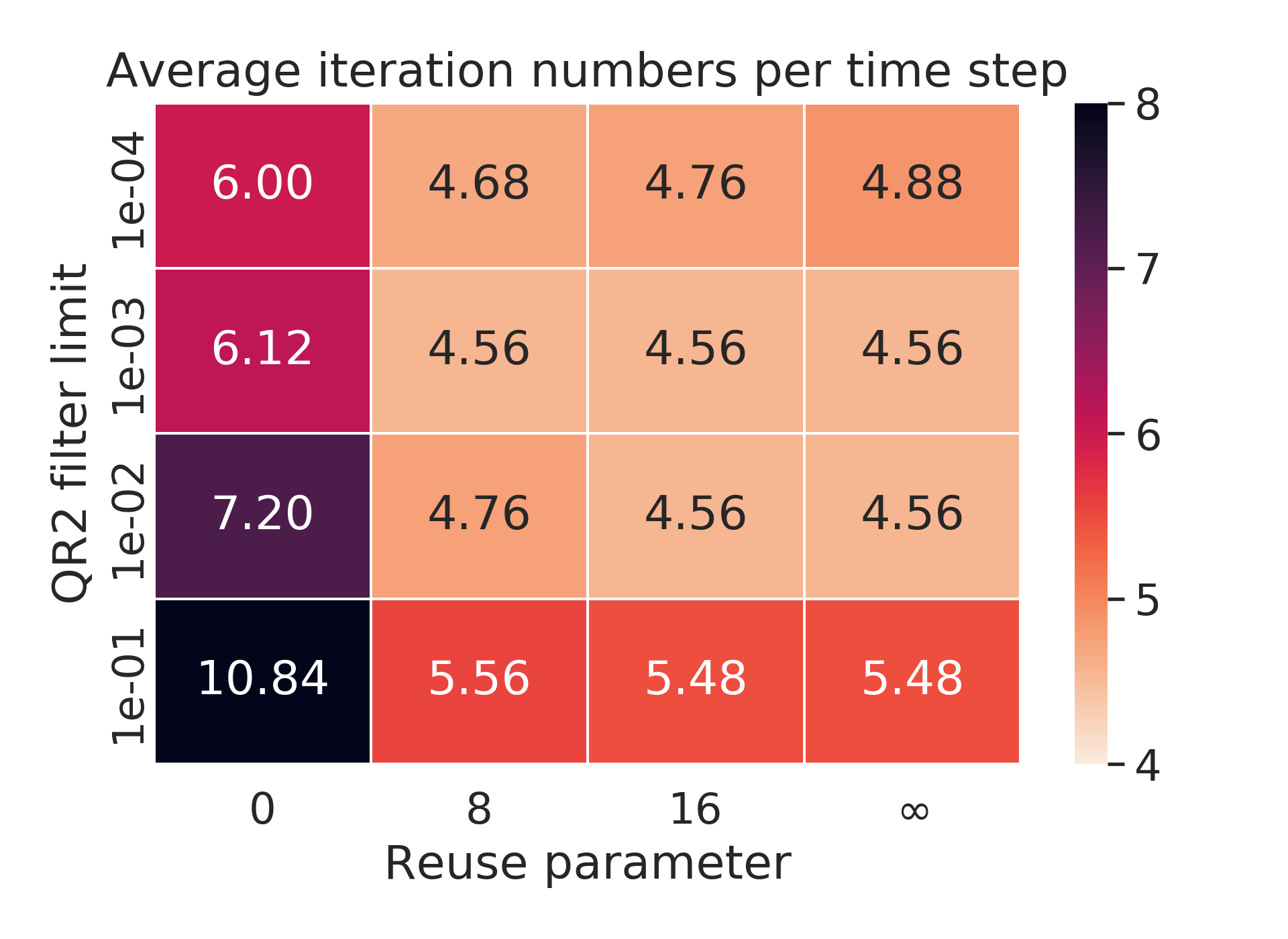}
    \caption{Using $\aperture$}
  \end{subfigure}
  ~
  \caption{Average number of coupling iterations per time step. The serial-implicit coupling and the IQN-ILS quasi-Newton method are used. The initial relaxation parameter is $\relaxationParameter = 0.1$, the filter limit and the reuse parameter are varied.}%
  \label{fig:newton-investigation:coupling-iter:ils}
  \begin{subfigure}[c]{0.47\textwidth}
    \includegraphics[keepaspectratio,width=1.1\linewidth]{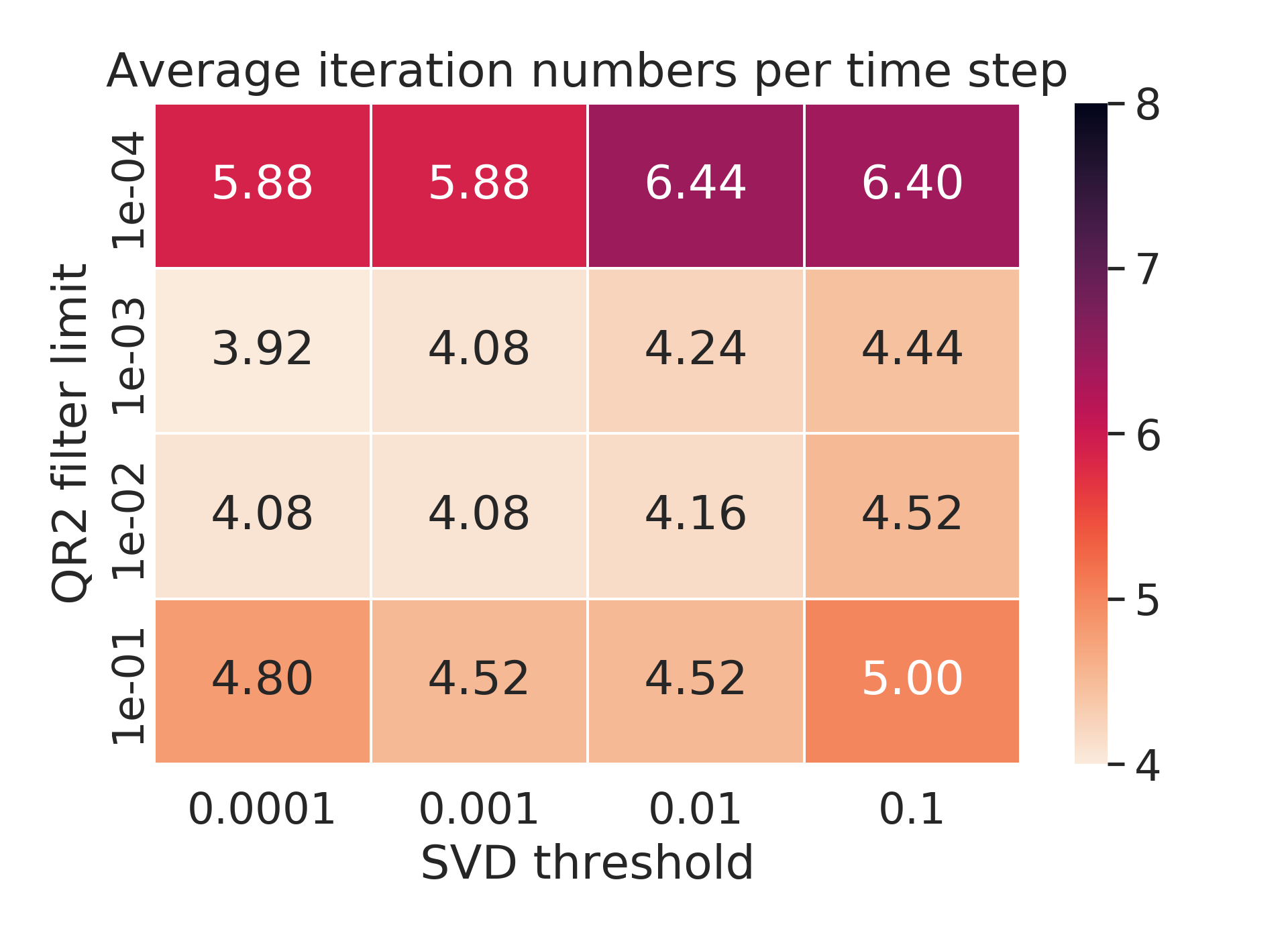}
    \caption{Using $\displacement^\pm$}
  \end{subfigure}
  ~
  \begin{subfigure}[c]{0.47\textwidth}
    \includegraphics[keepaspectratio,width=1.1\linewidth]{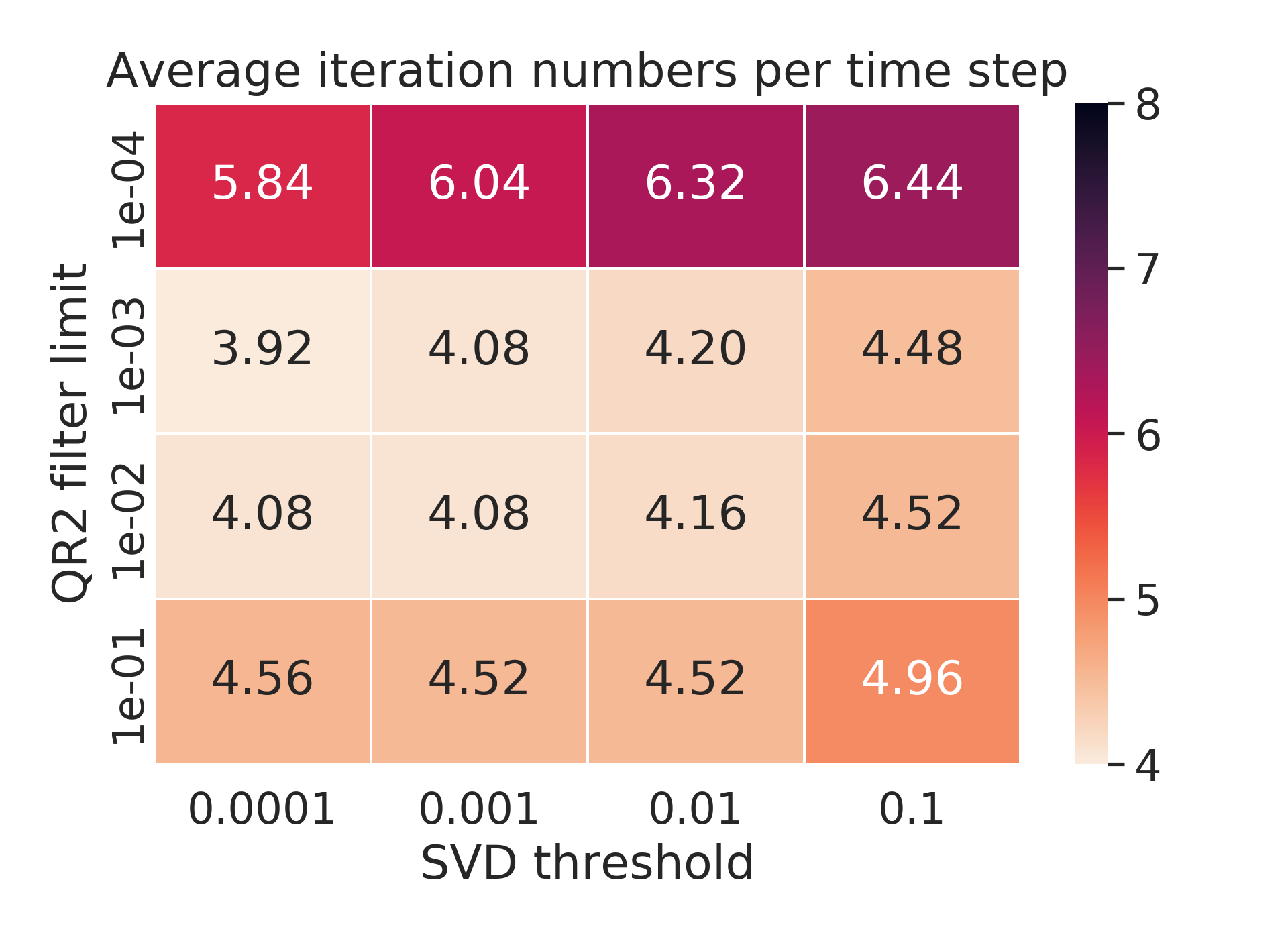}
    \caption{Using $\aperture$}
  \end{subfigure}
  \caption{Average number of coupling iterations per time step. The serial-implicit coupling and the IQN-IMVJ quasi-Newton method are used. The initial relaxation parameter is $\relaxationParameter = 0.1$, the filter limit and the reuse parameter are varied.}
  \label{fig:newton-investigation:coupling-iter:imvj}
\end{figure}

In \eqref{eq:fixed_point}, we have formulated the coupling in terms of the aperture $\aperture$. We refer to this as \enquote{pre-accumulated} case here as it uses $\aperture = \displacement^{+} \normalVector^+ + \displacement^{-} \normalVector^- $ in the coupling. An alternative way, that we have studied is the explicit use of the displacements $\displacement^\pm$ instead. We refer to this as \enquote{post-accumulated} approach. In the latter case, summation of the displacements to obtain the aperture is done in the fluid solver and the amount of data has to be exchanged between the solvers is increased. The latter case was straightforward to couple in \preCICE{} as it could rely on already implemented features. We do not expect, that the different approaches affect the obtained solution, but rather influence the coupling iteration convergence. Minimal deviations can be expected as the definition of the coupling residual differs slightly as it depends on $\aperture$ in the pre-accumulated case and on $\displacement^\pm$ otherwise. 

In Figs.~\ref{fig:newton-investigation:coupling-iter:ils} and \ref{fig:newton-investigation:coupling-iter:imvj}, the average number of coupling iterations per time step using the serial-implicit approach is given for the IQN-ILS and the IQN-IMVJ quasi-Newton method, while the filter limit and the quasi-Newton method-specific parameters are varied. The coupling works for all parameter settings investigated. 

The IQN-ILS method, see Fig.~\ref{fig:newton-investigation:coupling-iter:ils}, requires the largest number of coupling iterations, when no information from previous time steps is kept, i.e., the reuse parameter is $0$, and the filter limit is too large, i.e., $10^{-1}$. Consequently, the largest time-averaged number of coupling iterations is observed, when the reuse parameter is $\timeReuseParameter=0$ and the filter limit is $\filterLimit=10^{-1}$. All other cases need an average of approximately $4.5$ coupling iterations per time step, which is reasonably small.

For the IQN-IMVJ method, see Fig.~\ref{fig:newton-investigation:coupling-iter:imvj}, we observe only weak dependence on the SVD truncation threshold  $\svdTrunctationThreshold$, but stronger dependence on the filter limit. In contrast to the IQN-ILS method, the average number of iterations increases for larger ($\filterLimit=10^{-1}$)  and smaller filter limits ($\filterLimit=10^{-4}$). Bad choices of coupling parameters lead to approximately $6$ coupling iterations per time step, while, in the ideal cases, only $4$ iterations per time step are needed. This is even less than for the best cases of the ILS method.

The effect of the coupling parameters is very similar to what is reported in \cite{Scheufele2017}, where a study of different quasi-Newton methods was carried out for fluid-structure interaction test cases. The authors also observed, that the IQN-ILS method with $\timeReuseParameter=0$ performs worst, and than it depends stronger on the actual choice of parameters that the IQN-IMVJ method. 

Using the pre- or post-accumulation approach has barely any effect in the current setting. In the observed time frame, the time-averaged number of coupling iterations are nearly identical. However, using the aperture $\aperture$ in the coupling instead of the displacements $\displacement^{\pm}$ reduces the amount of data, that needs to be exchanged. Additionally, we observed strongly improved coupling stability for the fracture networks using the pre-accumulated coupling approach and, thus, it has been used in all following test cases.

\begin{figure}
  \centering
  \begin{subfigure}[c]{0.47\textwidth}
    \centering
    \includegraphics[keepaspectratio,width=1.1\linewidth]{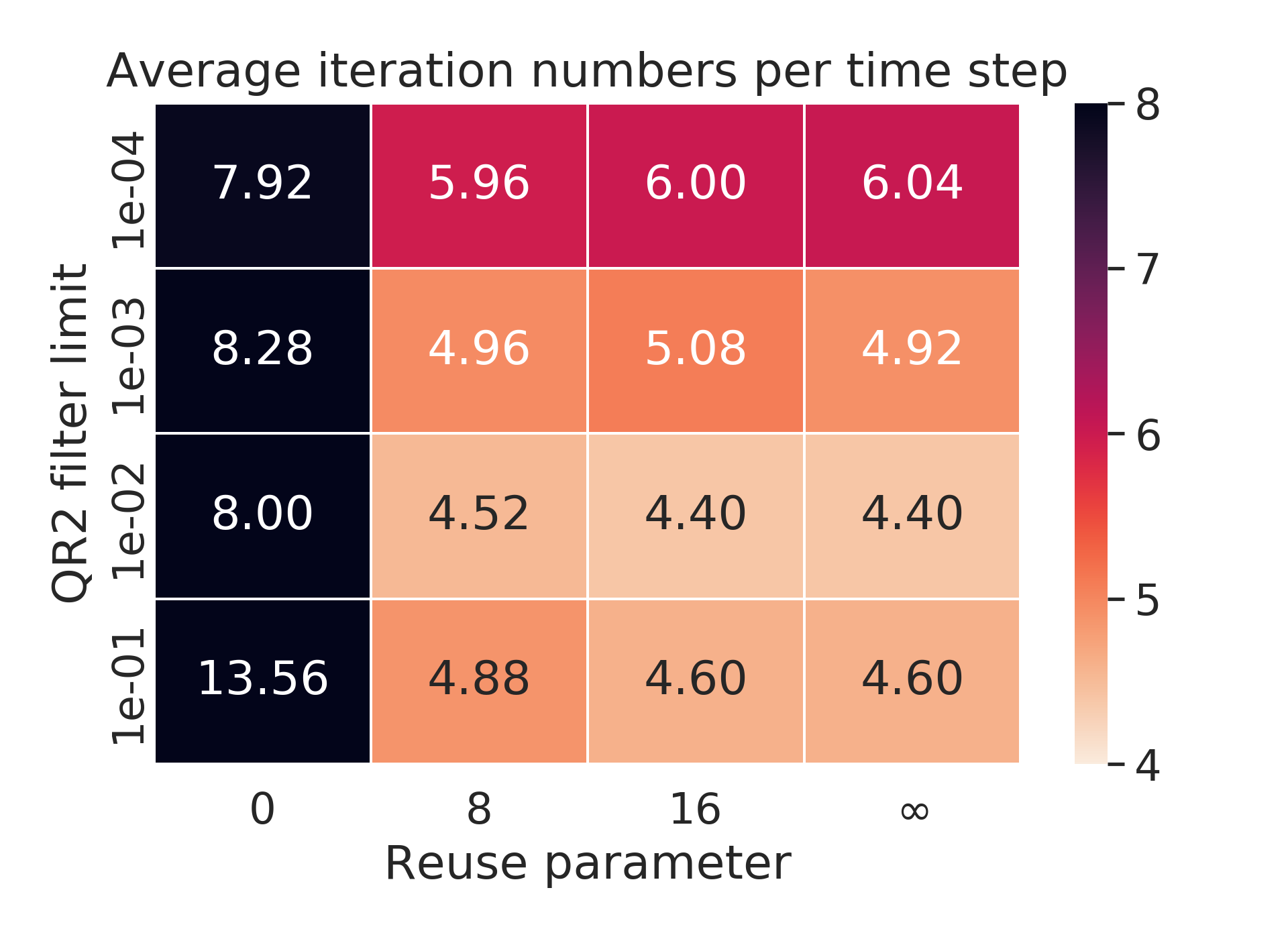}
    \caption{IQN-ILS}
  \end{subfigure}
  ~
  \begin{subfigure}[c]{0.47\textwidth}
    \includegraphics[keepaspectratio,width=1.1\linewidth]{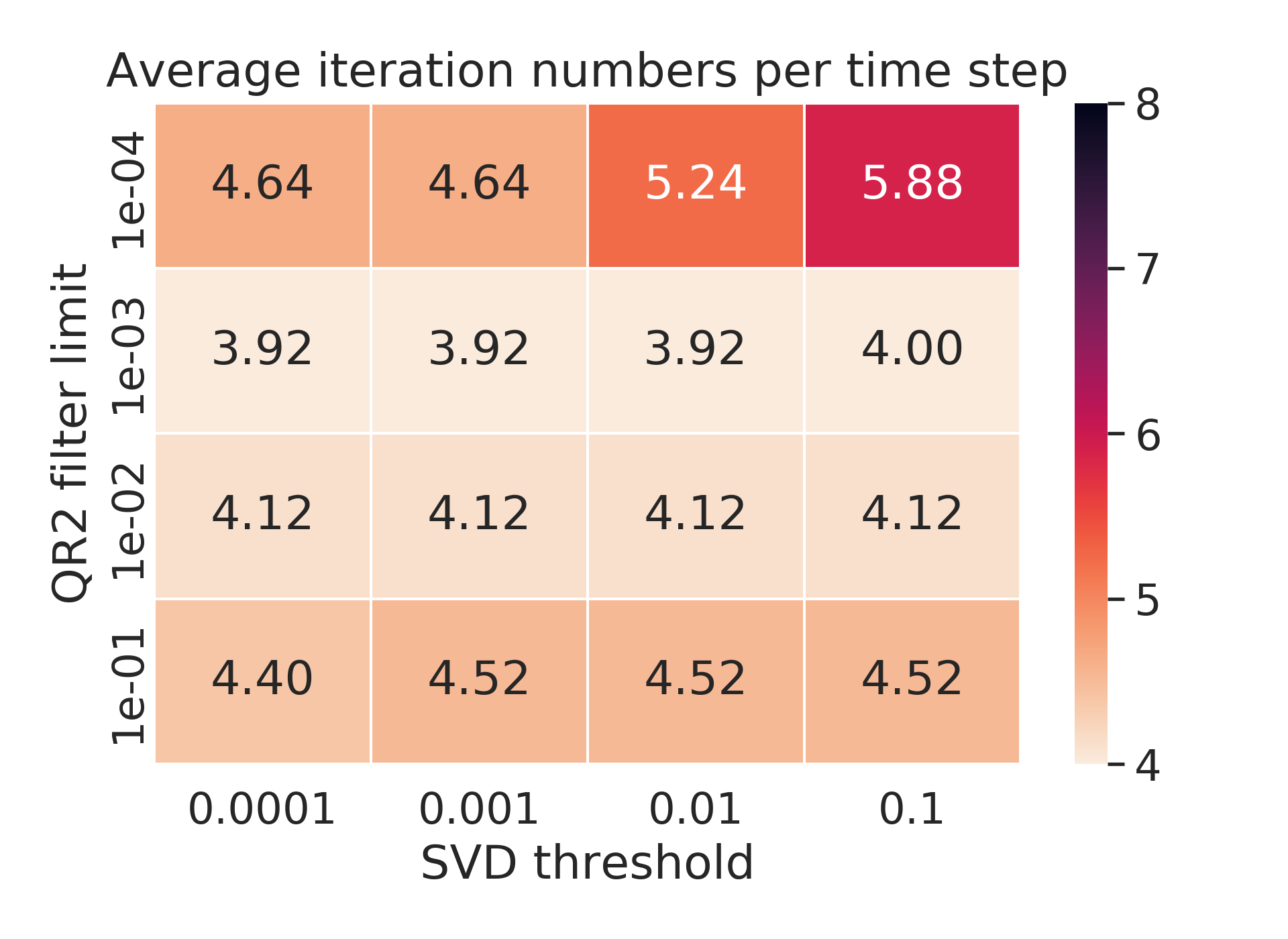}
    \caption{IQN-IMVJ}
  \end{subfigure}
  \caption{Average number of coupling iterations per time step. The parallel-implicit coupling and both quasi-Newton methods are used. The initial relaxation parameter is $\relaxationParameter = 0.1$ and the filter limit and quasi-Newton specific parameters are varied.}%
  \label{fig:newton-investigation:coupling-iter:parallel-implicit}
\end{figure}

In Fig.~\ref{fig:newton-investigation:coupling-iter:parallel-implicit}, we show the average number of coupling iterations per time step when a parallel-implicit coupling method is used. The coupling behavior is really similar to the serial-implicit approach, see Figs.~\ref{fig:newton-investigation:coupling-iter:ils} and \ref{fig:newton-investigation:coupling-iter:imvj}. The IQN-IMVJ method needs less iterations than the IQN-ILS method and shows less dependency on the coupling parameters. For the IQN-ILS, one needs most coupling iterations, again, for very small filter limits and especially when $\timeReuseParameter=0$. Surprisingly, the parallel-implicit even beats the serial-implicit coupling in terms of coupling iterations needed. This was not expected as the parallel-implicit approach is weaker and thus normally needs more iterations to recover the strong coupling behavior of the underlying physical problem. It is not clear what causes this and should be investigated further. 

\subsubsection{Mesh convergence study}
\label{sec:convergence:meshes}
For the numerical investigation of mesh convergence, we use three meshes for the fluid and the poro-elastic subdomain with different resolution: (i) ($68\,853$, $3\,292$) degrees of freedom (coarse, the same mesh as in the parameter study), (ii) ($447\,816$, $24\,188$) degrees of freedom (medium), and (iii) ($6\,586\,770$, $323\,976$) degrees of freedom (fine) in the porous matrix and fluid domain. The meshes are generated such that the location of the degrees of freedom in both domains match on the coupling interface, i.e.\, the grids are matching. The simulation setup is the same as before.  

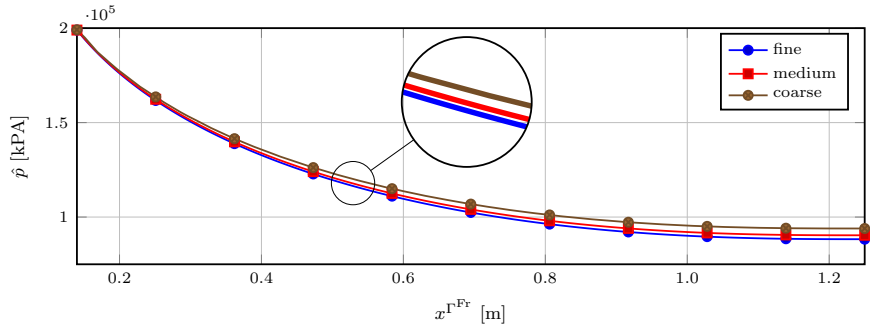
\begin{figure}[tb]
  \centering
    \centering
  \begingroup%
  \newcommand{\datapath}{.}%
      \begin{tikzpicture}[scale=1.,spy using outlines={circle, magnification=3, connect spies}]
    \begin{axis}[flowPressurePlotStyle, ymin=7.5e4,
    ymax=2e5,]
    \tikzset{>=latex}

    \addplot+[mark repeat={50}] table [x={"Points:0"}, y={"p"},col sep=comma] {\datapath/results/testcase-benchmark/nonlinear/non-aggl-si-ils-0_0045-dt-1e-1_QR2_0.01_reuse_8_relax_0.1/p_hd_flow_non_lin_2.5.csv};

    \addplot+[mark repeat={50}] table [x={"Points:0"}, y={"p"},col sep=comma] {\datapath/results/testcase-benchmark/nonlinear/non-aggl-si-ils-0_0175-dt-1e-1_QR2_0.01_reuse_8_relax_0.1/p_hd_flow_non_lin_2.5.csv};
    
    \addplot+[mark repeat={50}] table [x={"Points:0"}, y={"p"},col sep=comma] {\datapath/results/testcase-benchmark/nonlinear/non-aggl-si-ils-0_05-dt-1e-1_QR2_0.01_reuse_8_relax_0.1/p_hd_flow_non_lin_2.5.csv};

    \begin{scope}
    \spy[black,size=2cm] on (4.25,1.25) in node [fill=none] at (6,2.5);
    \end{scope}
    
    \legend{fine,  medium, coarse}
    \end{axis}
    \end{tikzpicture}%
  \endgroup
  \caption{Mesh convergence study for different including the comparison of different coupling approaches.}%
  \label{fig:meshstudy}
\end{figure}

In \figref{fig:meshstudy}, we present the pressure over the fracture radius at the final time $t=2.5$ for the different meshes. In all cases, a reasonable pressure curve is obtained. On the coarsest mesh, the pressure is highest. On the medium mesh, the pressure is clearly lower than on the coarse mesh. Thus, the predicted pressure tends to get lower for higher mesh resolution. This is confirmed by the solution on the finest mesh where the pressure is again lower than on the medium mesh. At the same time, the pressure difference between the fine and the medium mesh is smaller than between the medium and the coarse mesh although the difference in grid points increased severely. This indicates that the simulations converges toward a grid-converged solution.

The simulation setup was tested on shared-memory and distributed-memory systems with up 256 cores. No adjustments had to be made to the code as the parallelization is handled internally via \FEniCS and \preCICE. However, parallel efficiency is not the focus of this work, but the simplicity of the current approach to realize solvers and couplings that can be executed on parallel computers. Therefore, we do not report scaling results and instead leave it for future work.

\subsection{Injection and production in an arbitrarily fractured reservoir}
\label{sec:results_network}
\begin{table*}[tb]
\caption{Parameters defining the injection-production boundary value problem.}
\centering
\begin{adjustbox}{max width=\textwidth}
\begin{tabular}{llllll}\hline
\rule[1.9ex]{0ex}{1ex}\bf{Quantity} & \bf{Value} & \bf{Unit} & \bf{Quantity} & \bf{Value} & \bf{Unit} \\[1.1ex]\hline
\textbf{\textit{Numerical parameters}} &&&&&  \\
solid depth $\text{d}_{\poroElasticDomain_\text{Nw}}$ & $50.0$ & [m] & solid width $\text{w}_{\poroElasticDomain_\text{Nw}}$ & $50.0$ &  [m] \\
solid height $\text{h}_{\poroElasticDomain_\text{Nw}}$ & $50.0$ & [m] & small fracture radius $\text{r}^\text{sm}_{\fractureDomain_\text{Nw}}$ & $2.0$ &  [m] \\
large fracture radius $\text{r}^\text{la}_{\fractureDomain_\text{Nw}}$ & $4.5$ &  [m] &
solid DoF$_{\poroElasticDomain_\text{Nw}}$ & $9.5\cdot 10^{5}$ &  [-] \\
fracture DoF$_{\fractureDomain_\text{Nw}}$  & $5.0 \cdot 10^{4}$ & [-] & 
time step size $\Delta t$ & $5.0$ & [sec]\\
\textbf{\textit{Rock parameters}} &&&&&  \\
dry frame bulk modulus $K$ & $8.0$ & [GPa] & grain bulk modulus $K^\mathfrak{s}$ & $33.0$ & [GPa] \\
shear modulus $G$ & $15.0$ & [GPa] & initial porosity $\phi_0$ & $0.01$ & [-]  \\
intrinsic permeability $k^\mathfrak{s}$ & $1.1 \cdot 10^{-19}$  & [$\text{m}^2$] &
fluid compressibility $\beta^f$  & $0.45$  & [1/GPa] \\
effective bulk modulus $K_{eff}$ & $29.1$  & [GPa] &  effective shear modulus $G_{eff}$ & $15.4$  & [GPa] \\
\textbf{\textit{Fracture parameters}} &&&&&  \\
initial aperture $\delta_0$ & $75.0 $ & [$\mu$m] & effective fluid viscosity $\eta^{\mathfrak{f}R}$ & $0.001$ & [Pa$\cdot$s] \\
fluid compressibility $\beta^f$  & $0.45$  & [1/GPa] & injection pressure $\hat{p}_\text{i}^\text{Nw}$ & $200.0$ &  [kPa]\\
production pressure $\hat{p}_\text{p}^\text{Nw}$ & $-200.0$ &  [kPa] &&&\\
\textbf{\textit{Coupling Parameters}} &&&&&  \\
coupling & serial-implicit &  & quasi-Newton  & ILS &  \\
filter & QR2 & & initial relaxation $\omega$ & 0.001 &  [-]\\
rel. convergence & $10^{-5}$  & [-] &  & &  \\
\hline
\end{tabular}
\end{adjustbox}
\label{tab:fracture_network}
\end{table*}
\begin{figure}[tb]
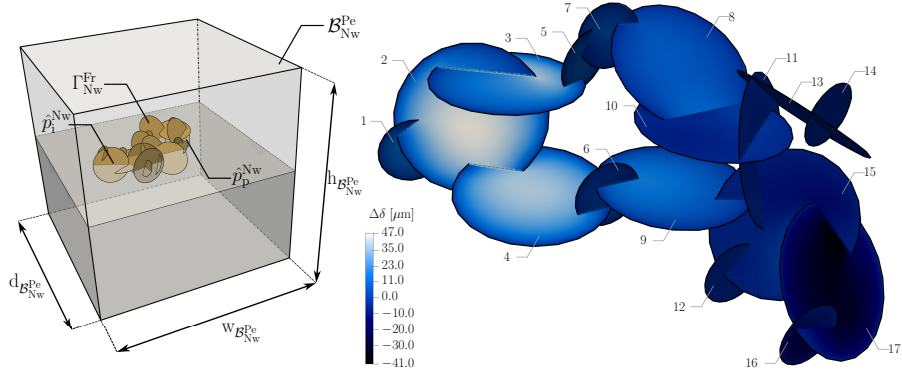

\begin{minipage}{0.4\textwidth}
\centering
    \resizebox{1.0\textwidth}{!}{\input{fig/numres/geometrical_set_up.eps_tex}}
   \end{minipage}\hfill
   \begin{minipage}{0.6\textwidth}
\centering
    \resizebox{1.0\textwidth}{!}{\input{fig/numres/aperture_network.eps_tex}}
   \end{minipage}
     \caption{\textbf{Left:} Sketch of the connected fracture network $\fractureDomain_\text{Nw}$ embedded in a solid matrix $\poroElasticDomain_\text{Nw}$. \textbf{Right:} Change of fracture aperture after $t=900$ secs including numbering of the embedded fractures.}\label{Fig:res_fracture_network_1}
\end{figure}

\begin{figure}[tb]
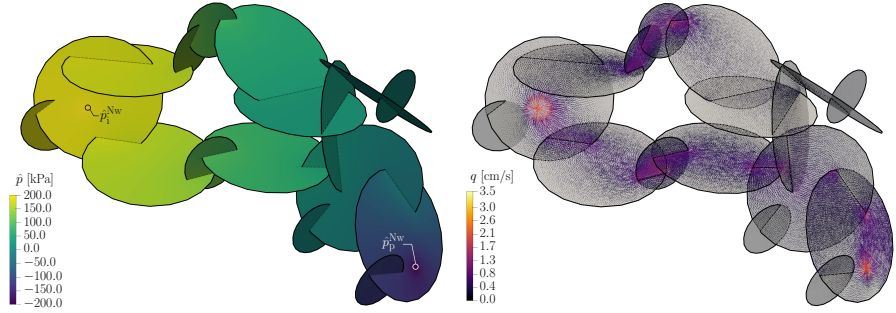

\begin{minipage}{0.5\textwidth}
\centering
    \resizebox{1.0\textwidth}{!}{\input{fig/numres/pressure_network.eps_tex}}
   \end{minipage}\hfill
   \begin{minipage}{0.5\textwidth}
\centering
    \resizebox{1.0\textwidth}{!}{\input{fig/numres/flow_glyphs.eps_tex}}
   \end{minipage}
     \caption{\textbf{Left:} Pressure state at time $t=900$ secs highlighting positions of pressure injection $\hat{p}_\text{i}^\text{NW}$ and production $\hat{p}_\text{p}^\text{NW}$. \textbf{Right:} Post-processed fluid flow field obtained by inserting pressure and aperture solutions into the balance of momentum \eqref{eq:hda_fluidflow} at time $t=900$ secs.}\label{Fig:res_fracture_network_2}
     \end{figure}
Transient flow and pressure data obtained by experimental field operations on fractured reservoirs provide information of their storage capacity, when evaluated by best numerical fits. Computational effort might be reduced for specific investigations on circular fractures by using two dimensional radial-symmetric models, but most field settings require consideration of several interacting fractures and three-dimensional modeling. 

Due to the low permeability of the surrounding bulk matrix, experimental pumping operations are often performed on fractured granite reservoirs. Such a problem setting might lead to instabilities throughout the numerical analysis, since characteristic pressure diffusion times of fracture network and granite bulk material greatly differ. Based on the short experimental execution time, outflow into the surrounding bulk material can be neglected and effective material parameters can be introduced based on Gassmann's solution defined by eq. (\ref{eq:gassmann}). This reduces the matrix response to linear-elastic behavior and limits the flow to the fracture domain. 

Here, we demonstrate the capability of the proposed method to solve flow problems in fracture networks embedded in a low permeable porous bulk material. Therefore, we test our approach on an arbitrarily generated fracture network containing 17 fractures by inducing injection and production Dirichlet pressure boundary conditions in the fracture domain $\fractureDomain_\text{PN}$, see \figref{Fig:res_fracture_network_1}. The regions of injection and production are highlighted. The parameters describing the boundary value problem and the coupling parameters are given in Tab.~\ref{tab:fracture_network}.

The applied coupling strategy shows a convergence behavior, which is characteristic for quasi-Newton schemes. Convergence in the first, critical time step is reached within 27 iterations, before the required number of iterations reduces to 14 in the second time step and reaches its minimum of 4--5 iterations for later time steps. The characterization of the tested network is carried out by investigations on the preferential flow path through the network connecting the injection with the production well. The chosen time step size resolves the fracture aperture evolution and allows to study transient hydro-mechanical effects such as the inverse pressure response at an early stage of the simulation. Nevertheless, the results are evaluated at time $t=900$ sec to focus on a solution close to the quasi-static equilibrium. 

In \figref{Fig:res_fracture_network_1}, the aperture changes of the fracture network indicate strong hydro-mechanical interaction showing, that fractures with dominant opening behavior are closing fractures with similar orientations by reallocation of the surrounding bulk material. This phenomenon is evident when looking at the aperture change distribution of fractures 9 and 10 and to some extend for fractures 1 and 2, or 15 and 17, respectively. 

The pressure distribution in \figref{Fig:res_fracture_network_2} shows a smooth pressure field, where pressure drops between fractures are highest, when large fractures are connected by small fractures which is due to their lower cross-section area and higher geometrical stiffness. The phenomenon is demonstrated best by the pressure drop between large fractures 3 and 8 interconnected by the small fractures 5 and 7. 

The calculated flow solution shown in \figref{Fig:res_fracture_network_2} visualizes the preferential flow path of the system through fractures 2, 4, 6, 9, 15 and 17 and confirms that the dominant opening of fracture 9 leads to the reduction of the flow through fracture 10. Regions of high flow rates are small connecting fractures and regions close to the injection or production area, which is consistent with the investigated pressure and aperture distributions. The study is representative for hydro-mechanical investigations on tested networks embedded in low permeable rock at a short time scale. It emphasizes the numerical capacity of the proposed approach and its relevance for detailed investigations of research questions in the field of experimental pumping operations.

\subsection{Flow through fractured porous media}
\label{sec:results_porous}
\begin{table*}[tb]
\caption{Parameters defining the fractured porous media boundary value problem.}
\centering
\begin{adjustbox}{max width=\textwidth}
\begin{tabular}{llllll}\hline
\rule[1.9ex]{0ex}{1ex}\bf{Quantity} & \bf{Value} & \bf{Unit} & \bf{Quantity} & \bf{Value} & \bf{Unit} \\[1.1ex]\hline
\textbf{\textit{Numerical parameters}} &&&&&  \\
solid depth $\text{d}_{\poroElasticDomain_\text{PN}}$ & $30.0$ & [m] & solid size $\text{w}_{\poroElasticDomain_\text{PN}}$ & $30.0$ &  [m] \\
solid height $\text{h}_{\poroElasticDomain_\text{PN}}$ & $20.0$ & [m] & smallest fracture are $\text{A}^\text{sm}_{\fractureDomain_\text{PN}}$ & $240.5$ &  [$\text{m}^2$] \\
largest fracture area $\text{A}^\text{la}_{\fractureDomain_\text{PN}}$ & $4.9$ &  [$\text{m}^2$] &
solid DoF$_{\poroElasticDomain_\text{PN}}$ & $6.0\cdot 10^{5}$ &  [-] \\
fracture DoF$_{\fractureDomain_\text{PN}}$  & $3.2 \cdot 10^{4}$ & [-] & 
time step width $\Delta t$ & $200.0$ & [sec]\\
\textbf{\textit{Rock parameters}} &&&&&  \\
dry frame bulk modulus $K$ & $8.0$ & [GPa] & grain bulk modulus $K^\mathfrak{s}$ & $33.0$ & [GPa] \\
shear modulus $G$ & $15.0$ & [GPa] & initial porosity $\phi_0$ & $0.01$ & [-]  \\
intrinsic permeability $k^\mathfrak{s}$ & $1.0 \cdot 10^{-17}$  & [$\text{m}^2$] &
fluid compressibility $\beta^f$  & $0.45$  & [1/GPa] \\
effective bulk modulus $K_{eff}$ & $29.1$  & [GPa] &  effective shear modulus $G_{eff}$ & $15.4$  & [GPa] \\
applied pressure $p_0^\text{PN}$ & $200.0$ &  [kPa] &
applied pressure $p_1^\text{PN}$ & $0.0$ &  [kPa] \\
\textbf{\textit{Fracture parameters}} &&&&&  \\
initial aperture $\delta_0$ & $75.0 $ & [$\mu$m] & effective fluid viscosity $\eta^{\mathfrak{f}R}$ & $0.001$ & [Pa$\cdot$s] \\
fluid compressibility $\beta^f$  & $0.45$  & [1/GPa] & &&\\
\textbf{\textit{Coupling Parameters}} &&&&&  \\
coupling & serial-implicit &  & quasi-Newton  & ILS &  \\
filter & QR2 & & initial relaxation $\omega$ & 0.001 &  [-]\\
rel. convergence & $10^{-5}$  & [-] &  & &  \\
\hline
\end{tabular}
\end{adjustbox}
\label{tab:porous_network}
\end{table*}

\begin{figure}[tb]
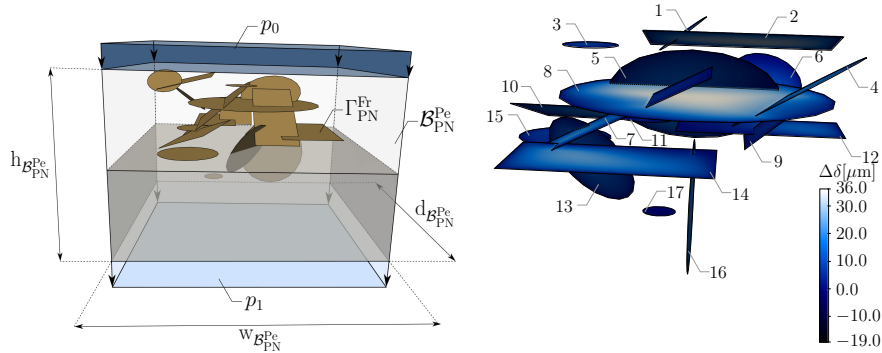

\begin{minipage}{0.5\textwidth}
\centering
    \resizebox{1.0\textwidth}{!}{\input{fig/numres/geometry_porous.eps_tex}}
   \end{minipage}\hfill
   \begin{minipage}{0.5\textwidth}
\centering
    \resizebox{1.0\textwidth}{!}{\input{fig/numres/aperture_porous.eps_tex}}
   \end{minipage}
     \caption{\textbf{Left:} Sketch of an arbitrarily chosen fracture network $\fractureDomain_\text{PN}$ embedded within a poro-elastic matrix highlighting the fluid pressure boundary conditions $p_0$ and $p_1$ applied to the poro-elastic domain $\poroElasticDomain_\text{PN}$. \textbf{Right:} Change of fracture aperture after $t=1.65$ days including numbering of the embedded fractures.}\label{Fig:res_porous_network_1}
\end{figure}

\begin{figure}[tb]
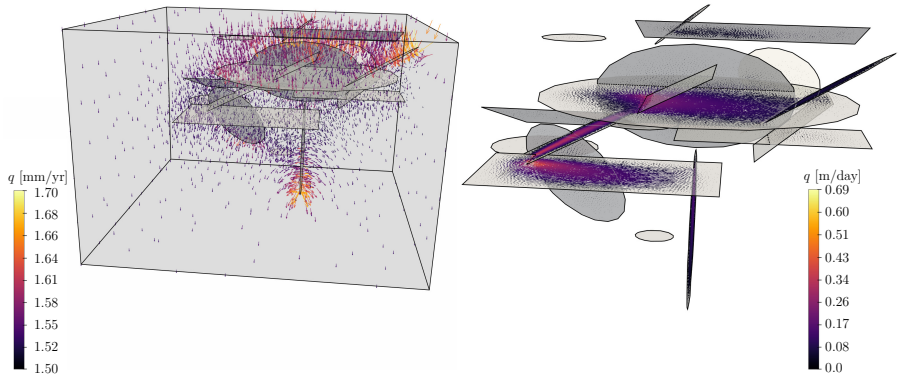

\begin{minipage}{0.50\textwidth}
\centering
    \resizebox{1.0\textwidth}{!}{\input{fig/numres/flow_porous.eps_tex}}
   \end{minipage}\hfill
   \begin{minipage}{0.50\textwidth}
\centering
    \resizebox{1.0\textwidth}{!}{\input{fig/numres/fracture_flow_porous.eps_tex}}
   \end{minipage}
     \caption{\textbf{Left:} Post-processed fluid-flow field in the poro-elastic domain $\poroElasticDomain_\text{PN}$ after $t=1.65$ days obtained by inserting the pressure solution into the governing equation \eqref{eq:poro_governing} and neglecting of time dependent terms by assuming quasi static conditions. \textbf{Right:} Post-processed fluid-flow field within the fracture domain $\fractureDomain_\text{PN}$ after $t=1.65$ days obtained by inserting the pressure and aperture solutions into the balance of momentum \eqref{eq:hda_fluidflow}.}\label{Fig:res_porous_network_2}\end{figure}
Approximations of preferential flow patterns through fractured poro-elastic media are of high interest in the field of nuclear waste disposal to reduce the risk of potential pollution by leak-off of contaminated matter. In contrast to the previous test case, an entirely different time scale is required, since investigation periods might last up to a million years. Discrete fracture networks influence the effective transport characteristics of a reservoir and even slight hydro-mechanically induced changes of the fractures' permeability have an immediate impact on its final characteristic diffusion time. 

The following boundary value problem investigates fractures embedded in a poro-elastic matrix with a low permeability, where a pressure gradient is induced on the poro-elastic domain by prescribing the fluid pressure $p_0$ and $p_1$ at the top and bottom, see \figref{Fig:res_porous_network_1}. 
The parameters defining the boundary value problem and the numerical coupling are given in Tab.~\ref{tab:porous_network}, the geometrical set up is shown in \figref{Fig:res_porous_network_1}. The embedded fractures can be grouped into three single fractures (fractures 3, 15 and 17), a small fracture network consisting of fractures 1 and 2 and a large fracture network formed by the remaining fractures (4--14 and 16). The position, shape and orientation of the fractures is chosen arbitrarily to demonstrate the flexibility of the method.

The applied coupling parameters result in a stable convergence behaviour, since convergence is reached within 31 iterations in the first, critical time step, 15 iterations in the following step and 4--5 iterations for later time steps. 

Investigations on the impact of hydraulically highly conductive deformable fractures on the transport characteristic of the tested reservoir are evaluated by means of the preferential flow pattern in the poro-elastic and the fracture domain. Results are displayed for a stage close to the converged quasi-static equilibrium after $t=1.65$ days, see Figs.~\ref{Fig:res_porous_network_1} and \ref{Fig:res_porous_network_2}. Similar to the findings of the previous study, hydro-mechanical interactions are evident in the fracture aperture change distribution displayed, where the fracture pairs 7 and 8, 9 and 14 and 4 and 8 have the strongest interactions, see \figref{Fig:res_porous_network_1}. The interaction results in opening or closing, respectively, of the involved fractures, which has an immediate impact on the local conductivity of the network. 

The post-processed flow solutions in the poro-elastic $\poroElasticDomain_\text{PN}$ and in the fracture domain $\fractureDomain_\text{PN}$ are presented in \figref{Fig:res_porous_network_2}. Depending on the fracture orientation and the fracture connectivity, embedded fractures have a distinct impact on the flow through the poro-elastic medium. Related to the orientation orthogonal to the pressure gradient and the lack of connectivity, the embedded single fractures 3, 15 and 17 have a minor contribution. At the same time, the fluid is strongly attracted by the small and large fracture network. The fluid mainly enters the networks through fractures 1, 4, 5 and 7 which are closest to the pressure boundary $p_0$ and is released back into the poro-elastic matrix through fractures 13 and 16 which are closest to the applied pressure boundary $p_1$. The impact of the fracture network on the transport characteristic of the studied poro-elastic domain is evident in terms of the distinct difference of flow rates in both domains. The hydro-mechanical interaction between the fracture and the poro-elastic domain has shown the potential of the method to consider flow through deformable fractures embedded in a hydro-mechanically interacting poro-elastic medium throughout long term investigations with a relevance for fields such as nuclear waste disposal. 

\section{Conclusion and outlook}
\label{sec:conclusion}
We proposed a new partitioned coupling approach for hydro-mechanical flow processes in deformable fractures embedded in a poro-elastic medium. Implicit coupling of the decomposed fracture flow and poro-elastic domain under consideration of the introduced interface conditions was realized by an iterative approach. The latter solves the underlying fixed-point problem using interface quasi-Newton methods. It was implemented using the open-source computing platform \FEniCS to solve the individual systems of PDEs and the open-source coupling library  \preCICE to realize the implicit coupling. 

We showed, that the proposed coupling strategy enables straight-forward usage of parallel computations throughout solution and coupling steps. This allows to study complex fracture systems in three dimensions with a high computational resolution. Evaluation of the proposed implementation against solutions obtained from a monolithic approach showed good agreement in terms of the transient pressure evolution in a single deformable fracture. Throughout a coupling convergence study, we showed the slightly better performance of the advanced IQN-IMVJ quasi-Newton scheme in comparison to the performance of the classical IQN-ILS quasi-Newton scheme, which is in good agreement with results on classical fluid-structure interaction problems in \cite{Scheufele2017}. 

The generality of the proposed strategy and its relevance for research topics such as modeling of injection and production in fractured reservoirs and the flow through a fractured poro-elastic domain was demonstrated throughout two numerical studies of complex fracture networks in three dimensions. We emphasized the advantage of the fracture and poro-elastic domain decomposition in terms of the creation of complex networks and straight forward post-processing of the numerical solutions in each computational region to identify preferential flow paths through the fracture network and the poro-elastic medium, respectively.

Future work can focus on the extension of the physical models to include temperature, e.g., extension of the partitioned coupling schemes, and the investigation of other discretization methods in for the subproblems. Considering additional physics in the model opens new application such as heat related energy production relevant for geothermal applications. The partitioned coupling schemes, especially in the black-box setting of \preCICE, can be improved by developing more sophisticated start-up strategies to reduce the number of coupling iterations in the first time steps and by new data mapping and communication concepts. The mixed-dimensional modeling leads to several challenges on how to communicate data between the different models, especially at fracture intersections, where the dominant deformation has to be identified.

\section*{Acknowledgments}
Holger Steeb and Patrick Schmidt gratefully acknowledge the funding provided by the German Federal Ministry of Education and Research (BMBF) for the GeomInt (I \& II) project (Grant Numbers 03A0004E and 03G0899E) in the BMBF Geoscientific Research Program “Geo:N Geosciences for Sustainability”. Alexander Jaust and Holger Steeb thank the DFG for supporting this work under Grant No.\ SFB 1313 (Project No.\ 327154368). We thank the \preCICE{} developers for their support, especially B.~Uekermann.

\printbibliography[]

\end{document}